\documentclass[
 superscriptaddress,
 amsmath,
 amssymb,
 aps,
 a4paper,
 floatfix,
 longbibliography,
 pra,
 twocolumn
]{revtex4-2}

\usepackage{graphicx}
\usepackage{dcolumn}
\usepackage{bm}

\usepackage{physics}

\usepackage{xcolor}

\usepackage{microtype}

\usepackage[american]{babel}
\usepackage[utf8x]{inputenc}

\usepackage{siunitx} 

\newcommand{\ee}{\text{e}}
\newcommand{\ii}{\text{i}}

\newcommand{\parabolicfunction}[2]{\mathcal{D}_{#1}(#2)}
\newcommand{\parabolicfunctiondot}[2]{\dot{\mathcal{D}}_{#1}(#2)}
\newcommand{\parabolicfunctionsquare}[2]{\mathcal{D}^{2}_{#1}(#2)}

\newcommand{\orcid}[1]{\href{https://orcid.org/#1}{\includegraphics[width=7pt]{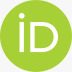}}}

\definecolor{cset-aps-blueberry}{RGB}{28,128,158}
\definecolor{cset-aps-blue}{RGB}{46,44,184}
\definecolor{cset-aps-turquoise}{RGB}{0,67,88}
\definecolor{cset-aps-limegreen}{RGB}{190,219,67}
\definecolor{cset-aps-green}{RGB}{31,138,112}
\definecolor{cset-aps-yellow}{RGB}{255,225,25}
\definecolor{cset-aps-orange}{RGB}{253,116,0}
\definecolor{cset-aps-red}{RGB}{219,0,43}
\definecolor{myred}{RGB}{255,0,20}

\usepackage{hyperref}
\hypersetup{%
    colorlinks=true,
    linkcolor={cset-aps-red},
    linkbordercolor={cset-aps-red},
    filecolor={cset-aps-orange},
    filebordercolor={cset-aps-orange},
    citecolor={cset-aps-blue},
    citebordercolor={cset-aps-blue},
    urlcolor={cset-aps-blueberry},
    urlbordercolor={cset-aps-blueberry},
    menucolor={cset-aps-limegreen},
    menubordercolor={cset-aps-limegreen},
    breaklinks=true,
    pdfborderstyle={/S/U/W 2},
    pdfpagemode=UseOutlines,
    pdfstartpage={1},
}

\usepackage{cleveref}

\AtBeginDocument{\RenewCommandCopy\qty\SI}

\begin{document}

\preprint{APS/123-QED}

\title{Hidden facts in Landau-Zener transitions revealed by the Riccati Equation}

\author{Eric P. Glasbrenner \orcid{0000-0002-0822-3888}}
\affiliation{Institut f{\"u}r Quantenphysik and Center for Integrated Quantum
    Science and Technology (IQST), Universit{\"a}t Ulm, Albert-Einstein-Allee 11, D-89081 Ulm, Germany}
\email{eric.glasbrenner@uni-ulm.de}
\author{Yannik Gerdes \orcid{0009-0005-0195-1259}}%
\affiliation{Institut f{\"u}r Quantenphysik and Center for Integrated Quantum
    Science and Technology (IQST), Universit{\"a}t Ulm, Albert-Einstein-Allee 11, D-89081 Ulm, Germany}
\email{yannik.gerdes@uni-ulm.de}
\author{Sándor Varró}
\affiliation{ELI-ALPS (Attosecond Light Pulse Source) Research Institute, ELI-HU Ltd., 6728 Szeged, Hungary}%
\author{Wolfgang P. Schleich \orcid{0000-0002-9693-8882}}
\affiliation{Institut f{\"u}r Quantenphysik and Center for Integrated Quantum
    Science and Technology (IQST), Universit{\"a}t Ulm, Albert-Einstein-Allee 11, D-89081 Ulm, Germany}%
\affiliation{Institute for Quantum Science and Engineering (IQSE), and Texas A\&M AgriLife Research and Hagler Institute for Advanced Study, Texas A\&M University, College Station, TX 77843-4242, USA}

\date{\today}

\begin{abstract}
We express the dynamics of the two probability amplitudes in the elementary Landau-Zener problem in terms of the solution of the corresponding Riccati differential equation and identify three key features: (i) The solution of the Riccati equation provides the bridge between the two probability amplitudes. (ii) Neglecting the non-linearity in the Riccati equation is equivalent to the Markov approximation which yields the exact asymptotic expression for one of the probability amplitudes, and (iii) the Riccati equation identifies the origin of the failure of the Markov approximation not being able to provide us in general with the correct asymptotic expression of the other probability amplitude. Our approach relies on approximate yet analytical solutions of the Riccati equation in different time regimes, highlighting the impact of its non-linear nature on the time evolution of the system.   
\end{abstract}

\maketitle

\section{Introduction}
Landau-Zener transitions \cite{Zener1932, Landau1932a, Landau1932b, Majorana1932, Stueckelberg1932} are prevalent in different branches of physics \cite{Hill1953, Peik1997, Konotop2005, Clade2009, SHEVCHENKO2010, Heim2013, Pagel2020, Gebbe2021, Kofman2023, kofman2024, konrad2024} and for this reason have been studied extensively \cite{Akulin1992, Garraway1992, Vitanov1999, Liu2002, Wubs2005, Wubs2006, rojo2010, Glasbrenner2023, lima2024}. In a recent article \cite{Glasbrenner2023}, we have presented a 'one-line' derivation of the exact asymptotic expression for one of the two probability amplitudes using the Markov approximation \cite{brewer1990, scullyzubairy1997, Paulisch2014}. In the present article, we show that both probability amplitudes of the Landau-Zener problem follow from a complex-valued Riccati equation \cite{Zheleznyakov1983, Gardas2010, bittanti2012, Schuch2014, Ndiaye2022} and discuss its solution. This approach demonstrates that the Markov approximation corresponds to neglecting the non-linearity in the Riccati equation. 

\subsection{First order but non-linear versus second order but linear differential equations}
In its most elementary form the Landau-Zener problem follows from a time-dependent Schrödinger equation for the probability amplitudes of a two state system with a time-independent coupling but a linear variation of the energy. It is well known that such a system can be decoupled by differentiating one more time, leading us to a linear differential equation of second order corresponding to a complex-valued harmonic oscillator with a time-dependent frequency. The parabolic cylinder functions \cite{abramowitz1965} are the solutions of this equation and lead us to the well-known asymptotic results \cite{Zener1932, Landau1932a, Landau1932b, Majorana1932, Stueckelberg1932, ivakhnenko2023}.
\par
On the other hand, there is a well established connection between differential equations of this type and the Riccati equation. The main difference is that the Riccati equation is a differential equation of \textit{first} order whereas the oscillator equation is of \textit{second} order. Unfortunately, this simplification comes at a high price, verifying again the well-known conservation law of complexity. In contrast to the oscillator equation which is linear, the Riccati equation contains a quadratic non-linearity.  
\par 
Despite this complication, the Riccati equation of the Landau-Zener problem derived in our article uncovers facts hidden so far in earlier approaches. In particular, the Riccati equation puts the Markov approximation on a solid foundation and is a consequence of neglecting the non-linearity.
\par
Although this non-linearity is not important for the asymptotic value of one of the two probability amplitudes, it is crucial for obtaining the correct asymptotic expression of the other one. This fact follows from a non-linear integral equation which is equivalent to the Riccati equation. 
\par
Moreover, we show that the two probability amplitudes are proportional to each other. For this reason, they enjoy a common phase factor determined by one of the amplitudes.
\par
The function connecting the two amplitudes is the solution
of the Riccati equation. As a consequence, one of the two probability amplitudes contains an additional phase factor following from the Riccati equation. 

\subsection{Overview and structure}
Our article is organized as follows: In \cref{section:2}, we first summarize the Landau-Zener problem and then depict the time evolution of the two probability amplitudes as trajectories in the complex plane. Here, we show the dynamics in the Schrödinger as well as the interaction picture. We conclude by briefly presenting the asymptotic values of the two probability amplitudes.
\par
We dedicate \cref{section:3} to the derivation of the Riccati differential equation for the Landau-Zener problem and express the two probability amplitudes in terms of the solution of the Riccati equation. We then derive two non-linear \textit{integral} equations which are equivalent to the Riccati \textit{differential} equation. One of these integral equations allows us to 
establish the absolute values of the solution of the Riccati equation without involving the normalization condition following from the unitarity of the Schrödinger equation. We conclude this section by presenting a numerical solution corresponding to the desired initial condition. This solution determines the behavior of both probability amplitudes.
\par
In \cref{section:4}, we recall the Markov approximation in the Schrödinger and interaction picture and demonstrate
that neglecting the non-linearity in the Riccati equation is equivalent to the Markov approximation. We conclude this section by comparing and contrasting the exact solution of the Riccati equation to the Markov solution.
\par
We dedicate \cref{section:5} to a detailed discussion of the Markov solution. Here, we derive approximate but analytical expressions for the three different domains of time, that is large negative and positive times and at the time of the transition. In the case of large negative times, we show that the special choice of the initial condition at a finite but large negative time causes oscillations in the probability amplitudes even before the level crossing. They do not appear if we involve initial conditions at minus infinity. 
\par
For large positive times, we derive the connection formula which provides a bridge between the solutions for negative and positive times. It brings out in a natural way the Stueckelberg oscillations and allows us to obtain the exact Landau-Zener result for one of the probability amplitudes.
\par
However, it fails when applied to the remaining probability amplitude. The origin of this failure is the neglected non-linearity in the Riccati equation, and the fact that this probability amplitude does not involve an \textit{integral} of the solution of the Riccati equation but the solution itself.
\par
In \cref{section:6}, we derive approximate but analytical expressions for the solution of the Riccati equation in the same time domains as in \cref{section:5}. Here, we emphasize especially the role of the non-linearity making the solution different from the Markov solution.
\par
We conclude in \cref{section:7} by summarizing our main results and by providing a brief outlook.
\par
We dedicate \cref{appendix:A} to the derivation of the two transition probability amplitudes and the solution of the Riccati equation in terms of parabolic cylinder functions. This approach which is valid for arbitrary initial conditions also provides us with the asymptotic expressions for the specific initial conditions addressed in this article. 
\par
In \cref{appendix:B}, we recall the instantaneous eigenstates of the time-dependent Hamiltonian in order to explain the appearance of oscillations even for large negative times. We show that they are a consequence of a superposition the two eigenstates.

\section{Landau-Zener model}
\label{section:2}
In this section, we provide an introduction into the Landau-Zener problem, introduce dimensional variables, and discuss the initial conditions. We then derive the coupled differential equations governing the two probability amplitudes in the Schrödinger and the interaction picture, and present their dynamics as trajectories in the complex plane. Finally, we briefly outline the standard method for obtaining the Landau-Zener formulae using parabolic cylinder functions. For the detailed calculation, however, is refer to \cref{appendix:A} or to Ref. \cite{ivakhnenko2023}, which presents an alternative approach.

\subsection{Schrödinger picture}
\label{subsection:formulation:of:the:landau:zener:problem}
This section addresses the Landau-Zener problem in the Schrödinger picture, describing the time evolution of the probability amplitudes from $-\infty$ to $+\infty$. We present a method to handle these asymptotic boundaries and conclude by illustrating the resulting dynamics. This approach highlights the characteristic features of Landau-Zener transitions.

\subsubsection{Formulation of the problem}
We consider the Schrödinger equation
\begin{align}
    \ii\hbar \frac{\dd}{\dd t}
    \begin{pmatrix}
        a(t) \\
        b(t)
    \end{pmatrix} = \mathcal{H}(t)
    \begin{pmatrix}
        a(t) \\
        b(t)
    \end{pmatrix} 
    \label{eq:Schrödinger:equation}
\end{align}
governed by the time-dependent Hamiltonian
\begin{align}
    \mathcal{H}(t) \equiv \hbar
    \begin{pmatrix}
    -\alpha t & \Omega \\
    \Omega & \alpha t
    \end{pmatrix}
    \label{eq:Hamiltonian}
\end{align}
where $\alpha$ and $\Omega$ denote the chirp rate and the coupling constant.

When we introduce the dimensionless time $\tau \equiv \Omega\cdot t$ the probability amplitudes $a$ and $b$ follow from the set of two coupled differential equations
\begin{align}
    \label{eq:diff:equation:a}
    \ii\dot{a}(\tau) &= -\epsilon\tau a(\tau) + b(\tau)\\
    \ii\dot{b}(\tau) &= a(\tau) + \epsilon\tau b(\tau)
    \label{eq:diff:equation:b}
\end{align}
where the dot denotes the differentiation with respect to $\tau$. Here, we have defined the ratio $\epsilon \equiv \alpha/\Omega^{2}$ between the chirp $\alpha$ and the square of the coupling constant $\Omega$.
\par
The initial condition for the probability amplitude $a$ in this article is given by 
\begin{align}
    a(\tau = -\tau_{0}) = 1
    \label{eq:initial:condition:a}
\end{align}
whereas the initial condition
\begin{align}
    b(\tau =  -\tau_{0}) = 0
    \label{eq:initial:condition:b}
\end{align}
for the probability amplitude $b$ follows from the normalization relation
\begin{align}
    \abs{a(\tau)}^{2} + \abs{b(\tau)}^{2} = 1
    \label{eq:normalization:relation}
\end{align}
together with the initial condition, \cref{eq:initial:condition:a}, for the probability amplitude $a$. 
\par
Here, we have introduced the large positive time $\tau_{0}$, which enables a rigorous analysis of the propagation of the probability amplitudes $a$ and $b$ from $-\infty$ to $+\infty$. Specifically, all calculations are performed for a finite value of $\tau_{0}$, starting at $\tau = -\tau_{0}$ and ending at $\tau=\tau_{0}$. The asymptotic limit $\tau_{0} = \infty$ is then considered only at the end.
\par
In this way, we avoid in the differential equation, \cref{eq:diff:equation:a}, the divergent initial condition for $\dot{a}$ arising from \cref{eq:initial:condition:a} and \cref{eq:initial:condition:b}. 
Furthermore, this approach is aligned with our numerical analysis where we integrate the differential equations, \cref{eq:diff:equation:a} and \cref{eq:diff:equation:b}, from a large but finite time $-\tau_{0}$, in complete accordance with Ref. \cite{Guttieres2023}.
\par
We emphasize that this approach, which involves considering large but finite values of $\tau_{0}$ before taking the asymptotic limit, reflects any experimental realization of a Landau-Zener transition, where probability amplitudes only propagate over a finite time interval. Nonetheless, this method presents significant challenges from a mathematical perspective.
\par
For example, the probability amplitude $a$ satisfies the differential equation of first order \cref{eq:diff:equation:a}. Therefore, the time dependence of the solution
\begin{align}
    a = a(\tau; a_{0})
    \label{eq:a:dependence:initial:condition:t:0}
\end{align}
depends crucially on the value $a_{0} \equiv a(-\tau_{0}, a_{0})$ of $a$ at the initial time $-\tau_{0}$ expressing the essence of the dependence of a differential equation on an initial condition.
\par
However, in our considerations the value $a_0$ is fixed to be unity but the initial time is changed. It is in this sense that the solution
\begin{align}
    a = a(\tau, -\tau_{0})
    \label{eq:a:dependence:initial:condition:t:0:1}
\end{align}
depends on the choice of the value of $\tau_{0}$ as indicated in \cref{eq:a:dependence:initial:condition:t:0:1} by the second argument. In order to distinguish the two dependencies expressed by \cref{eq:a:dependence:initial:condition:t:0} and \cref{eq:a:dependence:initial:condition:t:0:1} we have separated the time argument for the parameter by a semicolon and a comma, respectively. 
\par
We are interested in the so-defined value of $a$ at $\tau = \tau_{0}$, that is $a(+\tau_{0}; -\tau_{0})$, and its dependence on $\tau_{0}$. Obviously, for every choice of $\tau_{0}$, the solution $a$ is different and therefore also the final value. 

\subsubsection{Trajectories in the complex plane}
We are now in the position to numerically integrate \cref{eq:diff:equation:a} and \cref{eq:diff:equation:b} subjected to the initial conditions, \cref{eq:initial:condition:a} and \cref{eq:initial:condition:b}. Throughout this article, we utilize standardized libraries for scientific computing in all numerical simulations, such as SciPy \cite{Scipy2020} and Julia \cite{bezanson2017julia}.
\par
\begin{figure}[htbp]
    \includegraphics[width=\columnwidth]{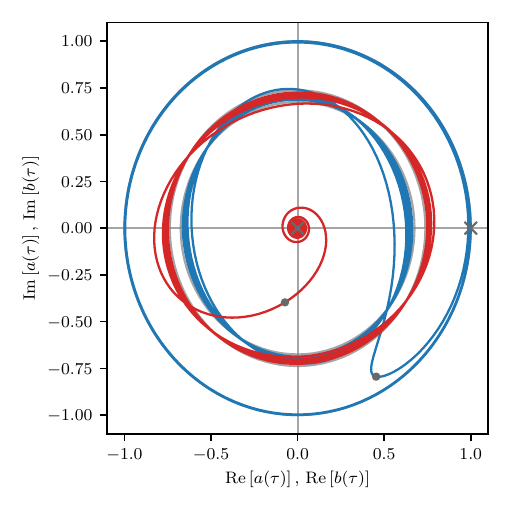}
    \centering
    \caption{Time dependence of the probability amplitudes $a$ (blue solid line) and $b$ (red solid line) of the Landau-Zener problem in the Schrödinger picture, depicted as trajectories in the complex plane. The grey crosses mark their starting points at the initial time $\tau =  -\tau_{0}$. The three barely visible grey circles represent: (i) the unit circle, (ii) the asymptotic value $\exp(-\pi/(2\epsilon))$ for $a$, and (iii) the asymptotic value $\sqrt{1-\exp(-\pi/\epsilon)}$ for $b$. The blue trajectory starts at the cross on the right-hand side, traverses the unit circle several times, and then jumps, in the vicinity of $\tau = 0$ (dark grey dot), to the circle determined by the asymptotic value. It also traverses this circle several times and performs oscillations around it. In the neighborhood of the jump, the trajectory reverses its velocity and turns around. For negative times $a$ rotates in the clockwise direction, whereas for positive times it moves counterclockwise. In contrast, the red trajectory begins at the origin of the complex plane and circumvents it with increasing amplitude several times before reaching its asymptotic value. It also performs oscillations around it. Again, the dark grey dot marks the time $\tau = 0$. For the probability amplitude $b$ the evolution is always in the clockwise direction. The trajectories shown for the parameters $\epsilon = 4.0$ and $\tau_{0} = 858.0855$ cover the time period $-\tau_{0}\leq\tau\leq\tau_{0}$.}
    \label{fig:5}
\end{figure}

In \cref{fig:5}, we present the dynamics of the probability amplitudes $a$ (blue solid line) and $b$ (red solid line) in the complex plane, obtained under the specified initial conditions, \cref{eq:initial:condition:a} and \cref{eq:initial:condition:b}, marked by grey crosses. Initially, the probability amplitude $a$ follows a circular trajectory with unit radius but begins to deviate slightly as time increases. Near $\tau = 0$ the trajectory reverses direction, transitioning from the circle of unit radius to a circle with radius $\exp(-\pi/(2\epsilon))$. Notably, the evolution proceeds clockwise for negative times and counterclockwise for positive times.
\par
On the other hand, the probability amplitude for $b$ starts to oscillate around the origin of the complex plane, forming a Cornu spiral, before transitioning near $\tau = 0$ to a circle trajectory with the radius $\sqrt{1-\exp(-\pi/\epsilon)}$. In contrast to $a$ the evolution  of $b$ remains consistently in the clockwise direction. The deviations of the probability amplitudes $a$ and $b$ from their final radii apparent in \cref{fig:5} are due to the Stueckelberg oscillations \cite{Stueckelberg1932}.  
\par
\begin{figure}[htbp]
    \includegraphics[width=\columnwidth]{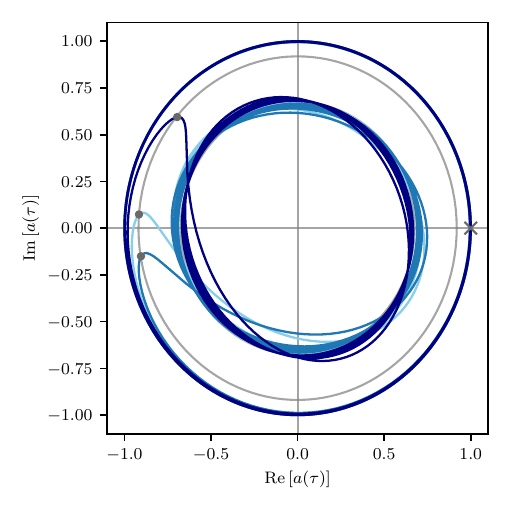}
    \centering
    \caption{Dependence of the probability amplitude $a$ on the choice of the initial time $-\tau_{0}$. The three trajectories corresponding to $\tau_{0}=8.581$ (light blue solid line), $\tau_{0}=27.135$ (blue solid line), and $\tau_{0}=54.270$ (dark blue solid line) all start at the same point in the complex plane, indicated by the grey cross. All trajectories eventually converge to the same final radius given by the asymptotic Landau-Zener formula $\exp(-\pi/(2\epsilon))$. However, the phases of $a$ at $\tau = 0$ (dark grey dots) are different for the three trajectories but not their absolute values as indicated by the grey circle passing through the dots. We emphasize that the $2\pi$-periodicity of the phase factors leads to a non-monotonic behavior of the point of jump $\tau = 0$. Indeed, the light blue line, associated with the smallest value of $\tau_{0}$, is surrounded by the curves corresponding to larger values of $\tau_{0}$. Here, we have chosen the parameter $\epsilon = 4.0$, and the trajectories presented cover the time period $-\tau_{0}\leq\tau\leq\tau_{0}$.}
    \label{fig:8}
\end{figure}

In order to bring out most clearly, the dependence of the probability amplitudes on the beginning of time, that is, on $\tau = -\tau_{0}$, we show in \cref{fig:8} the trajectories for $a$ for three different choices of $\tau_{0}$. As expected, all trajectories begin at the same point on the real axis of the complex plane, marked by the grey cross. However, their transitions around $\tau = 0$ occur at different points in the complex plane. Despite these differences, all trajectories ultimately converge to the same final radius, given by the asymptotic Landau-Zener result $\exp(-\pi/(2\epsilon))$.

\subsection{Interaction picture}
In a recent article \cite{Glasbrenner2023}, we have analyzed the Landau-Zener problem in the interaction picture given by the unitary transformation
\begin{align}
    \label{eq:transformation:interaction:a}
    \Tilde{a}(\tau) &= \ee^{-\ii \epsilon \tau^2 / 2} a(\tau)\\
    \Tilde{b}(\tau) &= \ee^{\ii \epsilon \tau^2 / 2} b(\tau)
    \label{eq:transformation:interaction:b}
\end{align}
where the \textit{tilde} indicates the probability amplitudes in the interaction picture.
\par
Hence, we arrive at the coupled differential equations
\begin{align}
    \label{eq:ODE:interaction:a}
    \ii \dot{\Tilde{a}}&= \ee^{-\ii \epsilon \tau^2} \Tilde{b}\\
    \ii \dot{\Tilde{b}}&= \ee^{\ii \epsilon \tau^2} \Tilde{a}
    \label{eq:ODE:interaction:b}
\end{align}
in the interaction picture with the corresponding initial conditions
\begin{align}
    \Tilde{a}(\tau = - \tau_0) = \ee^{-\ii \epsilon \tau_0^2 / 2}
    \label{eq:initial:condition:interaction:a}
\end{align}
and
\begin{align}
    \Tilde{b}(\tau = -\tau_0) = 0.
    \label{eq:initial:condition:interaction:b}
\end{align}
\par
At this point, we emphasize two characteristic differences to the corresponding equations in Ref. \cite{Glasbrenner2023}: (i) The quadratic phases in front of $\tilde{a}$ and $\tilde{b}$ differ by a factor of 2 which was lost in Ref. \cite{Glasbrenner2023} in the transition from the Hamiltonian, eq. 1 to the interaction picture given by eqs. 3 and 4. The replacement $\alpha\rightarrow 2\alpha$ in these and the following equations of Ref. \cite{Glasbrenner2023} makes them consistent with the Hamiltonian. (ii) In Ref. \cite{Glasbrenner2023}, we had chosen the initial condition for $\Tilde{a}(\tau = - \tau_0) = 1$ in contrast to \cref{eq:initial:condition:interaction:a} following from \cref{eq:initial:condition:a}. In order to obtain the same probability amplitudes as in Ref. \cite{Glasbrenner2023}, we have to choose
\begin{align}
    \tau_{0} \equiv \sqrt{\frac{2\pi}{\epsilon}T_{0}}
\end{align}
where $T_{0}$ is a large positive integer number.
\par
\begin{figure}[htbp]
    \includegraphics[width=\columnwidth]{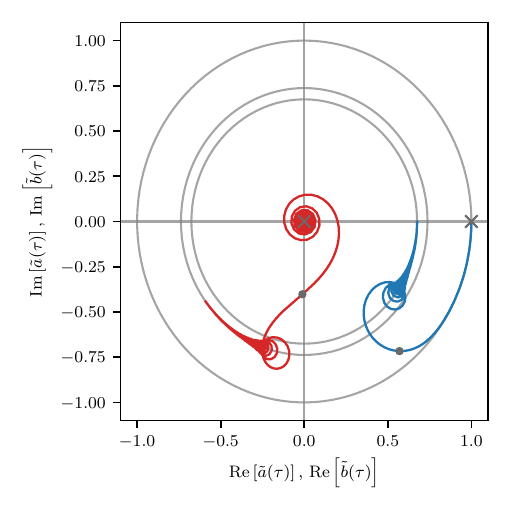}
    \centering
    \caption[]{Time dependence of the probability amplitudes $\tilde{a}$ (blue solid line) and $\tilde{b}$ (red solid line) of the Landau-Zener problem in the interaction picture depicted in the complex plane as trajectories. The grey crosses mark the starting points of the trajectories at the initial time $\tau =  -\tau_{0}$. The three grey circles, now clearly visible, represent: (i) the unit circle, (ii) the asymptotic value $\exp(-\pi/(2\epsilon))$ for $\tilde{a}$, and (iii) the asymptotic value $\sqrt{1-\exp(-\pi/\epsilon)}$ for  $\tilde{b}$. The blue trajectory starts at the cross on the right-hand side and, around $\tau = 0$ (dark grey dot), jumps to the circle determined by the asymptotic value without traversing the initial circle multiple times. In contrast, the red trajectory begins at the origin of the complex plane and circumvents it several times before reaching its asymptotic value. As in the case of the blue curve, the dark grey point represents $\tau = 0$. The decaying Stueckelberg oscillations emerging for positive times in both probability amplitudes are clearly visible. The trajectories presented for the parameter $\epsilon = 4.0$ and $\tau_{0} = 858.0855$ cover the time period $-\tau_{0}\leq\tau\leq\tau_{0}$.}
    \label{fig:6}
\end{figure}

In \cref{fig:6}, we present the dynamics of the probability amplitudes $\tilde{a}$ and $\tilde{b}$ in the complex plane where the respective initial conditions, \cref{eq:initial:condition:interaction:a} and \cref{eq:initial:condition:interaction:b}, are marked by grey crosses. Initially the probability amplitude for $\tilde{a}$ follows part of a circle of radius unity and close to $\tau = 0$ jumps to a circle with the radius $\exp(-\pi/(2\epsilon))$ and starts oscillating around this asymptotic value. The probability amplitude $\tilde{b}$ follows in the beginning a Cornu spiral. Close to $\tau = 0$ the probability amplitude for $\tilde{b}$ jumps to the final circle of radius $\sqrt{1-\exp(-\pi/\epsilon)}$ and also performs Stueckelberg oscillations \cite{Stueckelberg1932}.
\par
It is interesting to compare and contrast the two trajectories in the Schrödinger picture and the interaction picture depicted in \cref{fig:5} and \cref{fig:6}. Three characteristic features emerge: (i) The rotations around the origin in the complex plane prominent in the Schrödinger picture disappear in the interaction picture, (ii) the revolutions in the two pictures exchange their directions, and (iii) the decay of the Stueckelberg oscillations and the approach towards the asymptotic Landau-Zener result stands out most clearly in the interaction picture.  

\subsection{Linear second order differential equation}
One way to decouple the equations, eqs. (\ref{eq:diff:equation:a}) and (\ref{eq:diff:equation:b}), for the two probability amplitudes, is to differentiate the equation for $a$ once more, and use the equation for the derivative of $b$ which yields
\begin{align}
    \ddot{a}(\tau) + \left[\epsilon^{2}\tau^{2} + 1 - \ii\epsilon\right]a(\tau) =0
    \label{eq:differential:equation:second:order:a}
\end{align}
subjected to the initial conditions given by \cref{eq:initial:condition:a} and
\begin{align}
    \dot{a}(\tau = -\tau_{0}) = -\ii\epsilon\tau_{0}
    \label{eq:initial:condition:a:dot}
\end{align}
which follows from \cref{eq:diff:equation:a} in combination with initial condition, \cref{eq:initial:condition:b}, for $b$. 
\par
The solutions of \cref{eq:differential:equation:second:order:a} are well known \cite{abramowitz1965} to be the parabolic cylinder functions, provided \cref{eq:differential:equation:second:order:a} is first converted to the standard form 
\begin{align}
    \frac{\dd^{2}}{\dd z^2}a(z) - \left[\frac{z^{2}}{4} + c\right]a(z) = 0 
    \label{eq:standard:form:diff:equation:parabolic:cylinder:function}
\end{align}
where we have used the complex scaling
\begin{align}
     z \equiv \sqrt{2\epsilon}\tau\ee^{\ii\frac{\pi}{4}}
\end{align}
and introduced the constant
\begin{align}
    c \equiv \frac{\ii}{2\epsilon} + \frac{1}{2}.
\end{align}
\par
Indeed, \cref{eq:standard:form:diff:equation:parabolic:cylinder:function} is the basis for the derivation of the exact Landau-Zener formula. Two linearly independent solutions are needed to satisfy the initial conditions, \cref{eq:initial:condition:a} and \cref{eq:initial:condition:a:dot}. In \cref{appendix:A}, we show that in the limit of $\tau = \tau_{0}$ and $\tau_{0}\rightarrow\infty$ this combination provides us with the exact result \cite{Zener1932, Landau1932a, Landau1932b, Majorana1932, Stueckelberg1932, ivakhnenko2023, Glasbrenner2023}
\begin{align}
    a(\tau = \tau_{0} \rightarrow \infty) = \exp\left(-\frac{\pi}{2\epsilon}\right).
    \label{eq:landau:zener:formula:first}
\end{align}
Our derivation in \cref{appendix:A} suggests that this result is a consequence of a logarithmic phase singularity in combination with the well-known relation
\begin{align}
    -1 = \ee^{-\ii\pi}.
\end{align}
\par
Likewise, we find \cite{kazantsev1990, ivakhnenko2023, Wubs2005} in \cref{appendix:A} the asymptotic expression
\begin{align}
    \begin{split}
            b(\tau = \tau_{0} \rightarrow \infty) = \sqrt{1-\abs{a(\tau = \tau_{0} \rightarrow \infty)}^{2}}\,\ee^{\ii\phi}
            \label{eq:b:literature}
    \end{split}
\end{align}
with the phase
\begin{align}
    \begin{split}
            \phi \equiv \frac{3\pi}{4}  &- \epsilon\tau_{0}^{2} - \frac{1}{\epsilon}\ln\left(\sqrt{2\epsilon}\tau_{0}\right)\\ 
            &+ \arg\left[\Gamma\left(1+\frac{\ii}{2\epsilon}\right)\right]
            \label{eq:phi:eta:bar:infty}
    \end{split}
\end{align}
where $\Gamma$ denotes the Gamma function. 
\par
Indeed, our analysis brings out clearly that the asymptotic value of $b$ relies on the behavior of the parabolic cylinder function on both sides of the Stokes lines \cite{Kayanuma1997}. In contrast, the derivation of $a$ only requires this knowledge on one side.

\section{Riccati equation}
\label{section:3}
In this section, we express the two probability amplitudes $a$ and $b$ in terms of a solution of a non-linear differential equation of first order of the Riccati form. This equation will form the basis for elementary approximations employed to obtain in 'one-line' the Landau-Zener result, \cref{eq:landau:zener:formula:first}. In particular, we introduce \textit{two} different but equivalent expressions for the probability amplitude of $b$ in terms of the solution of the Riccati equation. This approach leads us to \textit{two} non-linear integral equations which are equivalent to the Riccati equation. We conclude this section by visualizing the dynamics of the Riccati equation by a numerical integration for a specific parameter.

\subsection{Derivation}
\label{sec:riccati:equation}
In order to derive the Riccati equation, we choose the ansatz
\begin{align}
    a(\tau) \equiv \ee^{-\ii\epsilon(\tau_{0}^{2}-\tau^{2})/2}\ee^{-\mathrm{H}(\tau)}
    \label{eq:ansatz:for:diff:eq:a}
\end{align}
with the abbreviation 
\begin{align}
    \mathrm{H}(\tau) \equiv \int\limits_{-\tau_{0}}^{\tau}\dd\tau^{\prime}\eta(\tau^{\prime})
    \label{eq:definition:H:ansatz:for:a}
\end{align}
for the integral of the function $\eta$ to be determined. We emphasize that the so defined probability amplitude $a$ satisfies the initial condition for $a$, \cref{eq:initial:condition:a}. 
\par
This ansatz leads us to the expression 
\begin{align}
    \dot{a} = \ii\epsilon\tau a - \eta a
     \label{eq:ansatz:for:diff:eq:a:first:derivative}
\end{align}
for the first derivative, and to 
\begin{align}
    \ddot{a} = a\left(\ii\epsilon - 2\ii\epsilon\tau\eta - \epsilon^{2}\tau^{2} - \dot{\eta} + \eta^{2}\right)
    \label{eq:ansatz:for:diff:eq:a:second:derivative}
\end{align}
for the second derivative given by \cref{eq:ansatz:for:diff:eq:a}.
\par
Next, we insert \cref{eq:ansatz:for:diff:eq:a} and \cref{eq:ansatz:for:diff:eq:a:second:derivative} into \cref{eq:differential:equation:second:order:a} and find 
\begin{align}
    0 = a \left(\eta^{2} - \dot{\eta} - 2\ii\epsilon\tau\eta + 1\right).
    \label{eq:insert:ansatz:into:diff:equation:riccati}
\end{align}
\par
Since $a = a(\tau)$ is non-vanishing, the right-hand side of the \cref{eq:insert:ansatz:into:diff:equation:riccati} must be zero. This condition leads us to the non-linear differential equation
\begin{align}
    \eta^{2} - \dot{\eta} - 2\ii\epsilon\tau\eta + 1 = 0
    \label{eq:riccati:equation:eta}
\end{align}
of first order of the Riccati form with the initial condition 
\begin{align}
    \eta(\tau =-\tau_{0}) = 0.
    \label{eq:initial:condition:for:eta}
\end{align}
\par
Obviously,
\begin{align}
    \eta \equiv \eta^{(R)} + \ii\eta^{(I)}
    \label{eq:real:imag:part:eta}
\end{align}
is a complex function with a real part $\eta^{(R)}\equiv\text{Re}[\eta]$ and an imaginary part $\eta^{(I)}\equiv\text{Im}[\eta]$.

\subsection{Probability amplitudes in terms of \texorpdfstring{$\eta$}{eta}}
In this section, we express the probability amplitudes $a$ and $b$ in terms of the solution $\eta$ of the Riccati equation. In particular, we show that that $b$ is determined by the product of $a$ and $\eta$. Here, the decomposition of $\eta$ into its real and imaginary part but also in its amplitude play a crucial role.

\subsubsection{Probability amplitude \texorpdfstring{$a$}{a}}
We start our analysis by considering the probability amplitude $a$ and substitute this representation, \cref{eq:real:imag:part:eta}, of $\eta$ into the ansatz, \cref{eq:ansatz:for:diff:eq:a}, and arrive at the identity
\begin{align}
    a(\tau) = \ee^{\ii\varphi(\tau)}A(\tau)
    \label{eq:ansatz:a:amplitude:and:phase}
\end{align}
where the amplitude
\begin{align}
    A(\tau) \equiv \ee^{-\Re\left[\mathrm{H}(\tau)\right]} = \exp\left(-\int\limits_{-\tau_{0}}^{\tau}\dd\tau^{\prime}\eta^{(R)}(\tau^{\prime})\right)
    \label{eq:definition:A}
\end{align}
is given by the negative exponential of the integral of the real part of $\eta$ and the phase
\begin{align}
    \varphi(\tau) \equiv -\frac{\epsilon}{2}\left(\tau_{0}^{2}-\tau^{2}\right) - \int\limits_{-\tau_{0}}^{\tau}\dd\tau^{\prime}\eta^{(I)}(\tau^{\prime})
    \label{eq:common:phase:varphi}
\end{align}
consists of a term increasing with $\tau^{2}$, and of the negative integrated imaginary part $\eta^{(I)}$ of $\eta$ given by $\Im\left[\mathrm{H}(\tau)\right]$. Hence, $\eta^{(I)}$ is a component of the phase velocity $\dot{\varphi}$. We note at this point that these expressions are similar to the one we obtained in \cite{Glasbrenner2024book}.
\subsubsection{Connection between \texorpdfstring{$a$}{a} and \texorpdfstring{$b$}{b} by \texorpdfstring{$\eta$}{eta}}
In order to find an expression for $b$ we rearrange \cref{eq:diff:equation:a} as follows
\begin{align}
   b = \ii\dot{a} + \epsilon\tau a,
\end{align}
and substitute \cref{eq:ansatz:for:diff:eq:a:first:derivative} into this expression. We arrive at the representation
\begin{align}
    b = -\ii\eta a
    \label{eq:expression:for:prob:amplitude:b:a}
\end{align}
for the probability amplitude for $b$ in terms of the complex-valued function $\eta$ and the probability amplitude $a$.
\par
Hence, $\eta$ given by the Riccati differential equation, \cref{eq:riccati:equation:eta}, determines \textit{both} probability amplitudes $a$ and $b$. Whereas $a$ is governed by the integral of $\eta^{(R)}$, $b$ is proportional to the product of $a$ and $\eta$. 
\par
It is interesting to also represent $b$ in terms of an amplitude and a phase. For this purpose, we  cast \cref{eq:expression:for:prob:amplitude:b:a} with the help of \cref{eq:ansatz:a:amplitude:and:phase} into the form
\begin{align}
    b(\tau) = \ee^{\ii\varphi(\tau)}\abs{\eta(\tau)}A(\tau)\ee^{\ii\psi(\tau)}
    \label{eq:representation:b:abs:eta:phase}
\end{align}
where we have introduced the definition 
\begin{align}
    \psi(\tau)\equiv \phi_{\eta}(\tau) - \frac{\pi}{2}
    \label{eq:definition:phase:psi}
\end{align}
governing the relative phase between the amplitudes $a$ and $b$ following from 
\begin{align}
    \eta(\tau) = \abs{\eta(\tau)}\ee^{\ii\phi_{\eta}(\tau)}.
    \label{eq:eta:abs:phase}
\end{align}
Due to the initial condition \cref{eq:initial:condition:for:eta} the phase $\phi_{\eta}$ of $\eta$ at $\tau = -\tau_{0}$ is not defined. For the sake of simplicity we choose in our calculations 
\begin{align}
    \phi_{\eta}(\tau=-\tau_{0}) \equiv \frac{\pi}{2}
    \label{eq:initial:condition:phi:eta}
\end{align}
leading us to the initial condition 
\begin{align}
    \psi(\tau=-\tau_{0}) = 0.
    \label{eq:initial:condition:psi}
\end{align}
\par
We note at this point that the phase $\varphi$ is a global phase which is the same for the probability amplitude $a$ and $b$.
\par
In addition, we substitute \cref{eq:expression:for:prob:amplitude:b:a} into the normalization condition, \cref{eq:normalization:relation}, and obtain the relation
\begin{align}
    \abs{a(\tau)}^{2} + \abs{a(\tau)}^{2}\abs{\eta(\tau)}^{2} = 1.
\end{align}
\par
When we solve for $\eta$, we find the expression 
\begin{align}
    \abs{\eta(\tau)} = \frac{\sqrt{1-\abs{a(\tau)}^{2}}}{\abs{a(\tau)}}
    \label{eq:abs:value:for:eta}
\end{align}
which together with the representation, \cref{eq:ansatz:a:amplitude:and:phase}, of $a$ yields
\begin{align}
    \abs{\eta(\tau)} = \frac{\sqrt{1-A(\tau)^{2}}}{A(\tau)}.
    \label{eq:abs:eta:1}
\end{align}
\par
As a result, we find from \cref{eq:eta:abs:phase} the expression
\begin{align}
    \eta(\tau) = \frac{\sqrt{1-A(\tau)^{2}}}{A(\tau)}\ee^{\ii\phi_{\eta}(\tau)}.
    \label{eq:different:representation:eta:A:phi}
\end{align}
for the solution $\eta$ of the Riccati equation.
\par
Moreover, when we substitute \cref{eq:abs:eta:1} into \cref{eq:representation:b:abs:eta:phase} we arrive at the formula
\begin{align}
    b(\tau) = \sqrt{1-A(\tau)^{2}}\ee^{\ii\varphi(\tau)}\ee^{\ii\psi(\tau)}
    \label{eq:probability:amplitude:A:varphi:psi}
\end{align}
for the probability amplitude $b$.
\par
\subsubsection{Summary}
In summary, we have shown that the two complex-valued probability amplitudes, $a$ and $b$, corresponding to four real-valued functions, are governed by the complex-valued Riccati differential equation, \cref{eq:riccati:equation:eta}, representing two real-valued functions. For the specific initial condition, \cref{eq:initial:condition:for:eta}, the time-integral of the real part $\eta^{(R)}$ and the phase $\phi_{\eta}$ of $\eta$ determine the absolute value $A$ of $a$ and the phase of $b$, respectively, given by \cref{eq:ansatz:a:amplitude:and:phase} and \cref{eq:probability:amplitude:A:varphi:psi}. Both amplitudes share a common time-dependent phase $\varphi$ defined by \cref{eq:common:phase:varphi} and are coupled by the normalization condition, \cref{eq:normalization:relation}.

\subsection{Non-linear integral equation}
In the preceding section, we have derived the Riccati differential equation, \cref{eq:riccati:equation:eta}, for $\eta$ which connects the two probability amplitudes. We devote the present section to obtain a non-linear integral equation for $\eta$ which allows us to establish two approximate solutions of the Riccati equation: (i) Neglecting the non-linearity leads us to the Markov solution which will be discussed in great detail in \cref{section:4}, and (ii) iterating the non-linear integral equation starting from the Markov solution yields the lowest approximation of the solution of the Riccati equation. 

\subsubsection{Exponential representation}
In the derivation of the expression, \cref{eq:expression:for:prob:amplitude:b:a}, for $b$, we have used the differential equation, \cref{eq:diff:equation:a}, along with the ansatz, \cref{eq:ansatz:for:diff:eq:a}, for $a$. Equivalently, we can formally integrate the differential equation, \cref{eq:diff:equation:b}, for $b$ subjected to the initial condition, \cref{eq:initial:condition:b}, to arrive at the formula 
\begin{align}
    b(\tau) = -\ii\ee^{-\ii\epsilon\tau^{2}/2}\int\limits_{-\tau_{0}}^{\tau}\dd\tau^{\prime}\ee^{\ii\epsilon\tau^{\prime 2}/2}a(\tau^{\prime}).
    \label{eq:formal:solution:for:b}
\end{align}
On first sight, \cref{eq:formal:solution:for:b} is very different from the representation of $b$ given by \cref{eq:expression:for:prob:amplitude:b:a} since now the probability amplitude $a$ appears under the integral, requiring knowledge of $a$ for all times prior to $\tau$. In contrast, in \cref{eq:expression:for:prob:amplitude:b:a}, $a$ is evaluated at $\tau$. 
\par
However, the exponential ansatz, \cref{eq:ansatz:for:diff:eq:a}, for $a$ and the elementary functional relation 
\begin{align}
    \ee^{\beta + \gamma} = \ee^{\beta}\ee^{\gamma}
\end{align}
for $c$-numbers $\beta$ and $\gamma$ provides a bridge between the two expressions.
\par
In order to bring out this fact most clearly, we first express $b$ given by \cref{eq:formal:solution:for:b} in the form 
\begin{align}
    b(\tau) = -\ii a(\tau)\ee^{-\ii\epsilon\tau^{2}/2}\int\limits_{-\tau_{0}}^{\tau}\dd\tau^{\prime}\frac{a(\tau^{\prime})}{a(\tau)}\ee^{\ii\epsilon\tau^{\prime 2}/2}.
\end{align}
\par
When we use the exponential ansatz, \cref{eq:ansatz:for:diff:eq:a}, for $a$ the relation
\begin{align}
\frac{a(\tau^{\prime})}{a(\tau)} = \ee^{-\ii\epsilon\tau^{2}/2}\ee^{\ii\epsilon\tau^{\prime 2}/2}\exp\left(-\int\limits_{\tau}^{\tau^{\prime}}\dd\tau^{\prime\prime}\eta(\tau^{\prime\prime})\right)
\end{align}
leads us to the representation 
\begin{align}
    \begin{split}
        b(\tau) = &-\ii a(\tau)\ee^{-\ii\epsilon\tau^{2}}\\ 
        &\times\int\limits_{-\tau_{0}}^{\tau}\dd\tau^{\prime}\exp\left(-\int\limits_{\tau}^{\tau^{\prime}}\dd\tau^{\prime\prime}\eta(\tau^{\prime\prime})\right)\ee^{\ii\epsilon\tau^{\prime 2}}.
    \end{split}
\end{align}
\par
Moreover, a comparison with \cref{eq:expression:for:prob:amplitude:b:a} yields the non-linear integral equation
\begin{align}
    \begin{split}
        \eta(\tau) &= \ee^{-\ii\epsilon\tau^{2}}\\
        &\times\int\limits_{-\tau_{0}}^{\tau}\dd\tau^{\prime}\exp\left(-\int\limits_{\tau}^{\tau^{\prime}}\dd\tau^{\prime\prime}\eta(\tau^{\prime\prime})\right)\ee^{\ii\epsilon\tau^{\prime 2}}.
        \label{eq:eta:non:linear:equation}
    \end{split}
\end{align}

We can easily verify the identity, \cref{eq:eta:non:linear:equation}, by differentiation which yields
\begin{align}
    \dot{\eta} = -2\ii\epsilon\tau\eta + 1 + \eta^{2},
    \label{eq:riccati:equation:different:representation}
\end{align}
in complete agreement with the Riccati equation, \cref{eq:riccati:equation:eta}. Moreover, the integral, \cref{eq:eta:non:linear:equation}, vanishes for $\tau = -\tau_{0}$ satisfying the initial condition, \cref{eq:initial:condition:for:eta}, for $\eta$. 
\subsubsection{Approximations of the non-linear integral equation}
It is interesting that $\eta^{2}$ appears in \cref{eq:riccati:equation:different:representation} due to the presence of $\eta$ in the exponent of \cref{eq:eta:non:linear:equation}. When we neglect $\eta$ in this exponent, that is the non-linearity in the Riccati equation, \cref{eq:riccati:equation:different:representation}, we arrive at the linear differential equation 
\begin{align}
    \dot{\eta}_{M} = -2\ii\epsilon\tau\eta_{M} + 1
    \label{eq:riccati:equation:different:representation:markov}
\end{align}
with the solution 
\begin{align}
    \eta_{M}(\tau) \equiv \ee^{-\ii\epsilon\tau^{2}}\int\limits_{-\tau_{0}}^{\tau}\dd\tau^{\prime}\ee^{\ii\epsilon\tau^{\prime 2}}.  \label{eq:eta:non:linear:equation:markov:solution}
\end{align}
Here, we have introduced the subscript $M$ since we will return in \cref{section:4} to this solution in the context of the Markov approximation.
\par
Moreover, since \cref{eq:eta:non:linear:equation} is an integral equation we are inclined to iterate the equation using the Markov solution, \cref{eq:eta:non:linear:equation:markov:solution}, as a starting point. By direct differentiation, we verify that the iterated approximate solution 
\begin{align}
    \begin{split}
            \eta_{I}(\tau) &\equiv \ee^{-\ii\epsilon\tau^{2}}\\
            &\times\int\limits_{-\tau_{0}}^{\tau}\dd\tau^{\prime}\exp\left(-\int\limits_{\tau}^{\tau^{\prime}}\dd\tau^{\prime\prime}\eta_{M}(\tau^{\prime\prime})\right)\ee^{\ii\epsilon\tau^{\prime 2}}
            \label{eq:eta:non:linear:equation:with:markov:solution}
    \end{split}
\end{align}
satisfies the differential equation 
\begin{align}
    \dot{\eta}_{I} = -2\ii\epsilon\tau\eta_{I} + 1 + \eta_{I}\eta_{M},
    \label{eq:riccati:equation:different:representation:with:markov}
\end{align}
where we have introduced the subscript $I$ to indicate that the solution emerged from an iteration. 
\par
When we compare the differential equations, \cref{eq:riccati:equation:different:representation} and \cref{eq:riccati:equation:different:representation:with:markov}, for $\eta$ and $\eta_{I}$, we note that the quadratic non-linearity of the Riccati equation is replaced by the product of $\eta_{I}$ and $\eta_{M}$. Hence, for $\eta_{I}\approx\eta_{M}$, we find $\eta_{I}\eta_{M} \approx \eta_{I}^2$ and $\eta_{I}$ is then an approximate solution of the Riccati equation, \cref{eq:riccati:equation:different:representation}. 

\subsection{Implicit solution of \texorpdfstring{$\eta$}{eta}}
With the help of the connection, \cref{eq:expression:for:prob:amplitude:b:a}, between the two probability amplitudes $a$ and $b$, and the normalization condition, \cref{eq:normalization:relation} we have been able to obtain the expression, \cref{eq:different:representation:eta:A:phi}, for $\eta$. Its absolute value given by \cref{eq:abs:eta:1} is determined by the amplitude $A$, \cref{eq:definition:A}, of $a$. We now cast the non-linear integral equation, \cref{eq:eta:non:linear:equation}, into a form which allows us to rederive \cref{eq:abs:eta:1} without using the normalization condition.

\subsubsection{A different integral equation for \texorpdfstring{$\eta$}{eta}}
In order to cast $\eta$ given by the non-linear integral equation, \cref{eq:eta:non:linear:equation}, into the form 
\begin{align}
    \eta = \ii\frac{b}{a}
    \label{eq:representation:of:eta:b:over:a}
\end{align}
following from \cref{eq:expression:for:prob:amplitude:b:a}, we take the term $a^{-1}(\tau)$ out of the integral which yields the expression
\begin{align}
    \begin{split}
        \eta(\tau) &= \ee^{-\ii\epsilon\tau^{2}}\exp\left(\,\int\limits_{-\tau_{0}}^{\tau}\dd\tau^{\prime}\eta(\tau^{\prime})\right)\\
        &\times\int\limits_{-\tau_{0}}^{\tau}\dd\tau^{\prime}\ee^{\ii\epsilon\tau^{\prime 2}}\exp\left(-\int\limits_{-\tau_{0}}^{\tau^{\prime}}\dd\tau^{\prime\prime}\eta(\tau^{\prime\prime})\right).
    \end{split}
\end{align}
\par
Next, we decompose $\eta$ into its real and imaginary part following \cref{eq:real:imag:part:eta}, and recall the relation, \cref{eq:definition:A}, for $A$ which yields the formula 
\begin{align}
    \eta(\tau) = A^{-1}(\tau)\ee^{-\ii\lambda(\tau)}\int\limits_{-\tau_{0}}^{\tau}\dd\tau^{\prime}\ee^{\ii\lambda(\tau^{\prime})}A(\tau^{\prime})
    \label{eq:representation:eta:different:integral:form}
\end{align}
where we have introduced the abbreviation 
\begin{align}
    \lambda(\tau) \equiv \epsilon\tau^{2} - \int\limits_{-\tau_{0}}^{\tau}\dd\tau^{\prime}\eta^{(I)}(\tau^{\prime}).
\end{align}
\par
On first sight, \cref{eq:representation:eta:different:integral:form} seems to be an explicit equation for $\eta$. However, this impression is misleading since $A$ involves the integral of the real part of $\eta$, and $\lambda$ contains, besides the quadratic time dependence, the integral of the imaginary part of $\eta$.
\par
Hence, \cref{eq:representation:eta:different:integral:form} is a rather complicated integral equation for $\eta$. Nevertheless, we can gain some insight as we show in the next section.

\subsubsection{An integral identity}
Indeed, we now use the non-linear integral equation, \cref{eq:representation:eta:different:integral:form}, to establish the identity 
\begin{align}
    \abs{\mathcal{I}(\tau)} = \sqrt{1-A^{2}(\tau)}
    \label{eq:abs:I:eta}
\end{align}
where we have introduced the integral
\begin{align}
    \mathcal{I}(\tau)\equiv\int\limits_{-\tau_{0}}^{\tau}\dd\tau^{\prime}\ee^{\ii\lambda(\tau^{\prime})}A(\tau^{\prime}).
    \label{eq:definition:integral:square:root:derivation:b}
\end{align}
\par
From the derivative 
\begin{align}
    \dv{\tau} \abs{\mathcal{I}}^{2} = \dv{\tau}\left(\mathcal{I}\mathcal{I}^{\ast}\right) = \dot{\mathcal{I}}\mathcal{I}^{\ast} + \mathcal{I}\dot{\mathcal{I}}^{\ast}
\end{align}
of the absolute value squared of $\mathcal{I}$, we find with the help of \cref{eq:definition:integral:square:root:derivation:b} the identity
\begin{align}
    \dv{\tau} \abs{\mathcal{I}(\tau)}^{2}&=2A^{2}(\tau)\Re\left[A^{-1}(\tau)\ee^{-\ii\lambda(\tau)}\mathcal{I}(\tau)\right]
\end{align}
which with \cref{eq:representation:eta:different:integral:form} reduces to
\begin{align}
\begin{split}
    \dv{\tau} \abs{\mathcal{I}(\tau)}^{2}=2A^{2}(\tau)\eta^{(R)}(\tau).
    \label{eq:derivation:of:the:square:root:term:left:hand:side}
\end{split}
\end{align}
\par
By differentiation we find the relation
\begin{align}
    \dv{\tau}\left[1-A^{2}(\tau)\right] = -2A(\tau)\dot{A}(\tau) = 2A^{2}(\tau)\eta^{(R)}(\tau)
    \label{eq:derivation:of:square:root:term:right:hand:side}
\end{align}
where in the last step we have recalled the definition, \cref{eq:definition:A}, of $A$.
\par
After we have equated \cref{eq:derivation:of:the:square:root:term:left:hand:side} and \cref{eq:derivation:of:square:root:term:right:hand:side} we arrive at the relation
\begin{align}
    \dv{\tau}\abs{\mathcal{I}(\tau)}^{2} = \dv{\tau}\left[1-A^{2}(\tau)\right].
    \label{eq:relation:between:I:and:1:minus:A:square}
\end{align}
\par
Next, we integrate both sides of \cref{eq:relation:between:I:and:1:minus:A:square} from $-\tau_{0}$ to $\tau$ and note the initial conditions
\begin{align}
    \mathcal{I}(-\tau_{0}) = 0
\end{align}
following from the definition of the integral $\mathcal{I}$ given by \cref{eq:definition:integral:square:root:derivation:b},
and the condition 
\begin{align}
    A(-\tau_{0}) = 1
\end{align}
determined from \cref{eq:definition:A}.
\par
Thus, we arrive indeed at the identity \cref{eq:abs:I:eta}. We emphasize that in this derivation we have not employed the normalization condition, \cref{eq:normalization:relation}.

\subsubsection{Summary}
We are now in the position to make the connection between the non-linear integral equation, \cref{eq:representation:eta:different:integral:form}, for $\eta$
and the representation \cref{eq:different:representation:eta:A:phi}.
\par
Indeed, the identity, \cref{eq:abs:I:eta}, leads to the representation
\begin{align}
    \ee^{-\ii\lambda(\tau)}\int\limits_{-\tau_{0}}^{\tau}\dd\tau^{\prime}\ee^{\ii\lambda(\tau^{\prime})}A(\tau^{\prime}) = \sqrt{1-A(\tau)^{2}}\ee^{\ii\phi_{\eta}(\tau)}
    \label{eq:identity:for:prob:amplitude:b:lambda}
\end{align}
where the phase $\phi_{\eta}(\tau)$ is determined by the argument of the left-hand side of this equation, that is
\begin{align}
    \phi_{\eta} = -\lambda(\tau) + \arg\left[\,\int\limits_{-\tau_{0}}^{\tau}\dd\tau^{\prime}\ee^{\ii\lambda(\tau^{\prime})}A(\tau^{\prime})\right].
    \label{eq:exact:expression:for:phi:eta}
\end{align}
When we substitute \cref{eq:identity:for:prob:amplitude:b:lambda} into the non-linear integral equation, \cref{eq:representation:eta:different:integral:form}, we arrive at the expression \cref{eq:different:representation:eta:A:phi} for $\eta$. This result demonstrates that the asymptotic phase $\phi$ of $b$ given by \cref{eq:phi:eta:bar:infty} follows from an asymptotic evaluation of \cref{eq:exact:expression:for:phi:eta} for $\tau = \tau_{0}$. 

\subsection{Numerical solution}
\begin{figure}[htbp]
    \includegraphics[width=\columnwidth]{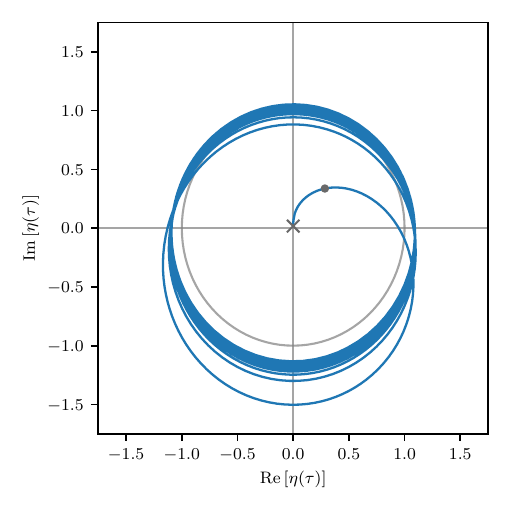}
    \centering
    \caption[]{Numerical solution $\eta$ of the Riccati differential equation, \cref{eq:riccati:equation:eta}, subjected to the initial condition \cref{eq:initial:condition:for:eta}, visualized as a trajectory in the complex plane. The trajectory starts at $\tau = -\tau_{0}$ from the origin marked by the grey cross and approaches rapidly without circling the origin an oscillatory regime. Indeed, the dark grey dot indicates the time $\tau = 0$. The grey unit circle serves as a guide for the eye and shows that this motion does not lie on the circle. The chosen parameters $\epsilon$ and $\tau_{0}$ are the same as those used in \cref{fig:5} and \cref{fig:6}.}
    \label{fig:7}
\end{figure}

In \cref{fig:7}, we present the numerical integration of the Riccati equation in the complex plane. The trajectory starts at the origin and quickly transitions into circular motion with oscillating amplitude. In contrast to \cref{fig:5}, the trajectory corresponding to $\eta$ does not circumvent the origin several times before reaching the asymptotic oscillatory regime. The grey dot indicating the point in time $\tau = 0$ shows that for negative times $\eta$ is almost zero and increases rapidly around $\tau = 0$ to a non-vanishing value.
\par
\begin{figure*}[htbp]
    \includegraphics[width=\textwidth]{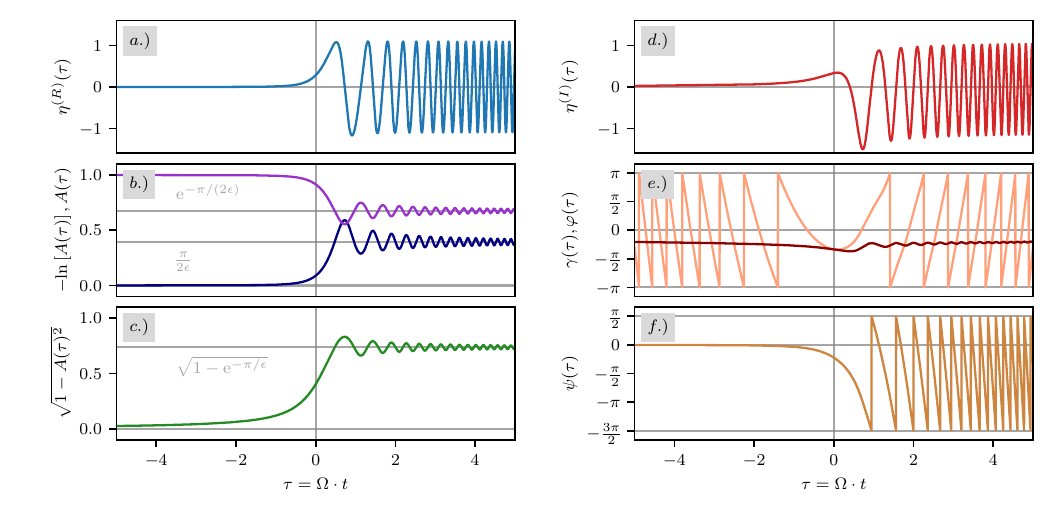}
    \centering
    \caption[]{Emergence of the time dependence of the two probability amplitudes $a$ and $b$ from the numerical solution $\eta$ of the Riccati differential equation, \cref{eq:riccati:equation:eta}, subjected to the initial condition, \cref{eq:initial:condition:for:eta}. The real part $\eta^{(R)}$ shown in $a.)$ by the blue solid line determines the absolute values of both probability amplitudes as indicated by the panels in the left column. Indeed, $b.)$ displays by the dark blue solid line the integrated real part of $\eta$. The  oscillations are a consequence of the oscillations of $\eta^{(R)}$ but their amplitude decreases due to the chirp in $\eta^{(R)}$ converging to the value $\pi/(2\epsilon)$ indicated by the lower grey horizontal line. The magenta curve represents the exponential of the dark blue solid line which corresponds to the absolute value $A$ of $a$, leading to the asymptotic Landau-Zener result $\exp\left(-\pi/(2\epsilon)\right)$ indicated by the upper grey horizontal line. In panel $c.)$, we depict the absolute value of $b$, obtained from the normalization condition \cref{eq:normalization:relation} and the time dependence of $A$.
    In panel $d.)$, we show by a red solid line the imaginary part $\eta^{(I)}$ which determines the phases of the two probability amplitudes $a$ and $b$. Indeed, both share a common phase $\varphi$ defined by \cref{eq:common:phase:varphi} and shown in panel $e.)$ by the light red solid line with the jumps which are the result of the $2\pi$-periodicity. This phase consists of the difference of a quadratic phase and the integral of the imaginary part of $\eta$, represented by the dark red curve. The phase of $\eta$, governed by $\eta^{(R)}$ and $\eta^{(I)}$, determines the phase $\psi$, shown in $f.)$ by the brown curve. Here, we have chosen the same parameters $\epsilon$ and $\tau_{0}$ as in \cref{fig:5}, \cref{fig:6} and \cref{fig:7}.}
    \label{fig:1}
\end{figure*}

 
This feature is most clearly observed in \cref{fig:1}, where panels $a.)$ and $b.)$ display the real part $\eta^{(R)}$ (blue solid line) and imaginary part $\eta^{(I)}$ (red solid line). We observe that $\eta^{(R)}$ vanishes for negative times but increases rapidly around $\tau = 0$ with a first maximum. Afterward, $\eta^{(R)}$ continues to oscillate between positive and negative values, with no significant decay in amplitude. Moreover, the frequency of these oscillations increases quadratically with time.
\par
In \cref{fig:1} b.) we depict $\eta^{(I)}$ as a function of time, starting from $\eta^{(I)}(\tau = -\tau_{0}) = 0$. As time increases, $\eta^{(I)}$ grows very slowly and reaches its maximum shortly after \mbox{$\tau = 0$}. For positive times, it oscillates, with the average of the oscillations gradually approaching the time axis. This behavior of $\eta^{(I)}$ contrasts with that of $\eta^{(R)}$ , as seen in \cref{fig:7}, where the trajectory in the complex plane slowly moves upward along the imaginary axis.
\par

We show the time integral of $\eta^{(R)}$ in \cref{fig:1} $c.)$, shown by the dark blue line below the plot of $\eta^{(R)}$. Additionally, we note that the time axes are identical in both figures. As $\eta^{(R)}$ increases from zero to its first maximum, the integral grows rapidly from zero to its first maximum as well. The oscillations in $\eta^{(R)}$ induce oscillations in the integral, known as Stueckelberg oscillations \cite{Stueckelberg1932}. Although, the oscillations in $\eta^{(R)}$ have the same amplitude, the oscillations of the integral gradually decay to the asymptotic value $\pi/(2\epsilon)$, as indicated by the horizontal grey line.
\par 
The exponential of the negative integral of the real part $\eta^{(R)}$ represented by the solid magenta line, determines the absolute value of $a$, i.e., the amplitude $A$. The second grey horizontal line indicates the Landau-Zener result, given by \cref{eq:landau:zener:formula:first}.
\par
From the normalization condition, \cref{eq:normalization:relation}, and the time dependence of $a$ we obtain the absolute value of $b$ shown in panel $e.)$ by the green curve. This value approaches the asymptotic limit of $b$, that is $\sqrt{1-\exp\left(-\pi/\epsilon\right)}$ indicated by the grey horizontal line.
\par
So far, we have addressed only the time dependence of the absolute values of $a$ and $b$ which arise exclusively from the time dependence of $\eta^{(R)}$, more precisely, from the time integral of $\eta^{(R)}$. We now focus on their phases, which will involve the real and imaginary part of $\eta$.
\par
According to \cref{eq:ansatz:a:amplitude:and:phase} and \cref{eq:probability:amplitude:A:varphi:psi} the amplitudes $a$ and $b$ share a common phase $\varphi$ defined by \cref{eq:common:phase:varphi}. This phase is shown in \cref{fig:1} $d.)$ it consists of the difference of a quadratic phase and 
the integral
\begin{align}
    \gamma(\tau) \equiv -\int\limits_{-\tau_{0}}^{\tau}\dd\tau^{\prime}\eta^{(I)}(\tau^{\prime})
\end{align}
of the imaginary part $\eta^{(I)}$. 
\par
The function $\gamma$ shown in \cref{fig:1} d.) by the dark red line, is negative and initially decreases slowly until it reaches a minimum at the first zero of $\eta^{(I)}$. This monotonic behavior is followed by a transition to an oscillatory regime, which eventually reaches the value $\varphi(\tau = \tau_{0})$. In complete analogy to $\eta^{(R)}$, the oscillations in $\gamma$, which give rise to the Stueckelberg oscillations, decay due to the increase of the frequency of the oscillations in $\eta^{(I)}$.
\par
The overall phase $\varphi$ shown in \cref{fig:1} d.) by the light red curve, spans several $2\pi$ intervals but varies slowly only in the neighborhood of $\tau = 0$. This behavior is consistent with \cref{fig:5}, where the trajectory corresponding to $a$ revolves around the origin multiple times on the unit circle before it moves to the inner circle and continues its rotations.
\par
Finally, the phase of $b$ is determined by the phase of $\eta$, shown by the brown curve in \cref{fig:1} $f.)$. In full agreement with the initial condition \cref{eq:initial:condition:psi}, $\psi = 0$ for $\tau=-\tau_{0}$ and remains zeros until $\tau = 0$ at which point it increases without bound. 
 
\section{Markov approximation and Riccati equation}
\label{section:4}

In this section, we explore the Markov approximation in both the Schrödinger and interaction pictures, comparing and contrasting the results in each. We then focus on the probability amplitude $a_{M}$ in Markov approximation and the exact amplitude $a$, which leads us to two different differential equations. By comparing them we identify the non-linearity in the Riccati equation for $\eta$ as the source of non-Markovian behavior and show that the Markov approximation is equivalent to neglecting this non-linearity. Finally, we compare numerical simulations of $\eta$ with those of $\eta_{M}$ and discuss their resulting differences. 

\subsection{Markov solution}
In the Schrödinger and interaction pictures, the two coupled differential equations for the probability amplitudes take different forms. As a result, the corresponding Markov approximations also differ. In order to bring this out most clearly, we perform the approximation in both frames in this section and compare the resulting expressions for $a$ and $\tilde{a}$.

\subsubsection{Schrödinger picture}
When we insert the formal solution, \cref{eq:formal:solution:for:b}, for $b$ into the differential equation, \cref{eq:diff:equation:a}, we obtain the differential equation
\begin{align}
    \dot{a}(\tau) = \ii\epsilon\tau a(\tau) - \ee^{-\ii\epsilon\tau^{2}/2}\int\limits_{-\tau_{0}}^{\tau}\dd\tau^{\prime}\ee^{\ii\epsilon\tau^{\prime 2}/2}a(\tau^{\prime})
    \label{eq:integro:diff:equation:for:a}
\end{align}
of Lippmann-Schwinger type \cite{Paulisch2014} for the probability amplitude $a$. Up to this point, our derivation remains exact, and no approximation has been introduced. 
\par
We now perform the Markov approximation
\begin{align}
    \ee^{-\ii\epsilon\tau^{\prime 2}/2}a(\tau^{\prime}) \approx \ee^{-\ii\epsilon\tau^{2}/2}a(\tau)
    \label{eq:markov:approximation}
\end{align}
which assumes that the main contribution of the probability amplitude $a$ to the integral in  \cref{eq:integro:diff:equation:for:a} arises from the upper limit, a condition that holds when $a(\tau)$ oscillates much more slowly than $\ee^{\ii\epsilon\tau^{2}}$. This case occurs precisely when the condition $1 \ll \epsilon\tau$ is satisfied.
\par
The Markov approximation yields the differential equation
\begin{align}
    \dot{a}_{M} = \left[\ii\epsilon\tau - \eta_{M}\right]a_{M}
    \label{eq:diff:equation:a:M}
\end{align}
where we have introduced the function
\begin{align}
    \eta_{M}(\tau) \equiv \ee^{-\ii\epsilon\tau^{2}}\int\limits_{-\tau_{0}}^{\tau}\dd\tau^{\prime}\ee^{\ii\epsilon\tau^{\prime 2}}
    \label{eq:definition:markov:function}
\end{align}
and the subscript $M$ to emphasize that the differential equation \cref{eq:diff:equation:a:M} is an approximation 
of the exact integro-differential equation \cref{eq:integro:diff:equation:for:a}, based on the Markov approximation, \cref{eq:markov:approximation}.
\par
It is straightforward to integrate \cref{eq:diff:equation:a:M} and we find the Markov solution
\begin{align}
    a_{M}(\tau) = \ee^{-\ii\epsilon\tau_{0}^{2}/2}\ee^{\ii\epsilon\tau^{2}/2}\ee^{-\mathrm{H}_{M}(\tau)}
    \label{eq:Markov:solution}
\end{align}
with the abbreviation
\begin{align}
    \mathrm{H}_{M}(\tau) \equiv \int\limits_{-\tau_{0}}^{\tau}\dd\tau^{\prime}\eta_{M}(\tau^{\prime}).
    \label{eq:H:M:definition}
\end{align}
Obviously, $a_{M}$ satisfies the initial condition \cref{eq:initial:condition:a}, that is 
\begin{align}
    a_{M}(-\tau_{0}) = 1.
    \label{eq:initial:condition:a:markov}
\end{align}. 
\par
In complete analogy to the exact expression, \cref{eq:ansatz:a:amplitude:and:phase}, for $a$, we can represent the Markov solution 
\begin{align}
    \begin{split}
        a_{M}(\tau) \equiv \ee^{\ii\varphi_{M}(\tau)}A_{M}(\tau)
    \end{split}
\end{align}
in amplitude
\begin{align}
    A_{M}(\tau) \equiv \ee^{-\Re\left[\mathrm{H}_{M}(\tau)\right]} = \exp\left(-\int\limits_{-\tau_{0}}^{\tau}\dd\tau^{\prime}\eta^{(R)}_{M}(\tau^{\prime})\right)
    \label{eq:definition:A:markov}
\end{align}
and phase
\begin{align}
    \varphi_{M}(\tau) \equiv -\frac{\epsilon}{2}\left(\tau_{0}^{2}-\tau^{2}\right) - \int\limits_{-\tau_{0}}^{\tau}\dd\tau^{\prime}\eta^{(I)}_{M}(\tau^{\prime}).
    \label{eq:common:phase:varphi:markov}
\end{align}
\par
Here, we have recalled the definition, \cref{eq:H:M:definition}, of $\mathrm{H}_{M}$ and have decomposed $\eta_{M}$ into its real and imaginary part, that is 
\begin{align}
    \eta_{M} \equiv \eta^{(R)}_{M} + \ii\eta^{(I)}_{M}.
    \label{eq:real:imag:part:eta:markov}
\end{align}
\par
Indeed, this expression suggests the phase velocity
\begin{align}
    \dot{\varphi}_{M}(\tau) \equiv \epsilon\tau - \eta^{(I)}_{M}(\tau)
    \label{eq:phase:velocity:markov}
\end{align}
of $a_{M}$.

\subsubsection{Interaction picture}
Next, we compare and contrast this expression with the one derived from the Markov approximation in the interaction picture, which was previously obtained in Ref. \cite{Glasbrenner2023}.
From \cref{eq:ODE:interaction:b}, we find with the initial condition, \cref{eq:initial:condition:interaction:b}, the expression for $\tilde{b}$ which when substituted into \cref{eq:ODE:interaction:a} yields the integral differential equation
\begin{align}
    \dot{\tilde{a}}(\tau) = - \ee^{-\ii\epsilon\tau^{2}}\int\limits_{-\tau_{0}}^{\tau}\dd\tau^{\prime}\ee^{\ii\epsilon\tau^{\prime 2}}\tilde{a}(\tau^{\prime})
    \label{eq:integro:diff:equation:for:a:interaction:picture}
\end{align}
in the interaction picture. 
\par
Due to the transformation, \cref{eq:transformation:interaction:a}, into the interaction picture, the Markov approximation, \cref{eq:markov:approximation}, in the Schrödinger picture takes the form
\begin{align}
    \tilde{a}(\tau^{\prime})\approx\tilde{a}(\tau)
    \label{eq:markov:approximation:interaction:picture}
\end{align}
and the approximated integral differential equation reads
\begin{align}
    \dot{\tilde{a}}_{M}(\tau) = - \eta_{M}(\tau)\tilde{a}_{M}(\tau)
    \label{eq:diff:equation:a:M:interaction:picture}
\end{align}
with the solution
\begin{align}
    \tilde{a}_{M}(\tau) =  \ee^{-\ii \epsilon \tau_{0}^{2} / 2}\ee^{-\mathrm{H}_{M}(\tau)}
    \label{eq:Markov:solution:old}
\end{align}
satisfying the initial condition \cref{eq:initial:condition:interaction:a}.
\par
A comparison between the expressions for $a_{M}$ and $\tilde{a}_{M}$ given by \cref{eq:Markov:solution} and \cref{eq:Markov:solution:old} establishes the relation
\begin{align}
    a_{M}(\tau) = \ee^{\ii\epsilon\tau^{2}/2}\tilde{a}_{M}(\tau),
    \label{eq:Markov:solution:relation:different:pcitures}
\end{align}
in complete agreement with the transformation \cref{eq:transformation:interaction:a} into the interaction picture.
\par
We conclude by noting that in Ref. \cite{Glasbrenner2023}, the initial condition was expressed in the interaction picture rather than in the Schrödinger picture. As a result, the quadratic phase factor in $\tau_{0}$ does not appear in the expression for $\tilde{a}_{M}$ in Ref \cite{Glasbrenner2023}. Moreover, we have already performed the limit $\tau_{0} \rightarrow \infty$ in the lower boundary of the integral in \cref{eq:Markov:solution:old}.

\subsection{Non-linearity as a source of non-Markovianity}
The function $\eta_{M}$, defined by \cref{eq:definition:markov:function}, and its role in determining the probability amplitude $a_{M}$ via the Markov solution, \cref{eq:Markov:solution}, are closely related to the function $\eta$ governed by the Riccati equation, \cref{eq:riccati:equation:eta}. To highlight this analogy most clearly, we now differentiate the definition of $\eta_{M}$, given in \cref{eq:definition:markov:function}, once more and obtain the \textit{linear} differential equation 
\begin{align}
    \dot{\eta}_{M} + 2\ii\epsilon\tau\eta_{M} - 1 = 0
    \label{eq:diff:equation:from:eta:M}
\end{align}
of first order. A comparison to the Riccati equation, \cref{eq:riccati:equation:eta}, reveals that the two differential equations differ by the non-linear term $\eta^{2}$.

In order to emphasize this point, we now rewrite the Riccati equation, \cref{eq:riccati:equation:eta}, in the form
\begin{align}
    \dot{\eta} + 2\ii\epsilon\tau\eta = 1 + \eta^{2}
    \label{eq:differential:equation:rearranged}
\end{align}
which allows us to translate it into an integral equation
\begin{align}
    \begin{split}
            \eta(\tau) &= \eta(\tau = -\tau_{0})\ee^{-\ii\epsilon\tau^{2}}\\
            &+ \ee^{-\ii\epsilon\tau^{2}}\int\limits_{-\tau_{0}}^{\tau}\dd\tau^{\prime}\left[1 + \eta(\tau^{\prime})^{2}\right]\ee^{\ii\epsilon\tau^{\prime 2}}
    \end{split}
\end{align}
which with the help of the initial condition, \cref{eq:initial:condition:for:eta}, reduces to
\begin{align}
    \eta(\tau) = \eta_{M}(\tau) + \ee^{-\ii\epsilon\tau^{2}}\int\limits_{-\tau_{0}}^{\tau}\dd\tau^{\prime}\eta(\tau^{\prime})^{2}\ee^{\ii\epsilon\tau^{\prime 2}}.
    \label{eq:integral:equation:riccati:equation}
\end{align}
Here, we have recalled the definition, \cref{eq:definition:markov:function} of $\eta_{M}$.
\par
Hence, the Markov approximation is equivalent to neglecting the non-linear term $\eta^{2}$. 

\subsection{Markov versus exact}
\begin{figure}[htbp]
    \includegraphics[width=\columnwidth]{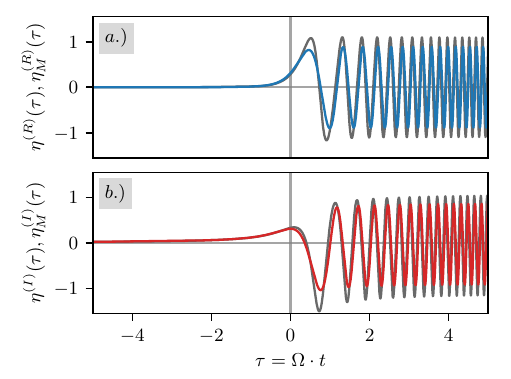}
    \centering
    \caption[]{Comparison between the exact numerical solution $\eta$ of the Riccati equation and the analytic Markov approximation $\eta_{M}$. Here, we depict the real $a.)$ and imaginary part $b.)$ of $\eta$ (solid grey line) and the approximated solution $\eta_{M}$ (colored solid line) which agree very well. For these numerical calculations we have chosen the parameter $\epsilon = 4.0$ and $\tau_{0} = 858.0855$.}
    \label{fig:3}
\end{figure}

In \cref{fig:3} we compare and contrast the functions $\eta$ and $\eta_{M}$ determining the probability amplitudes $a$ and $a_{M}$. It is amazing how well they agree qualitatively. In particular, $\eta_{M}$ reflects correctly the transition from an almost vanishing value to an oscillatory function. They only deviate in the amplitudes of their oscillations at positive times and their phases.  
We observe the largest deviation around the dimensionless time $\tau = 0$, which is due to the fact that at this point the Markov approximation is no longer valid, since the condition $1 \ll \epsilon\tau$ is not fulfilled. Nevertheless, it is remarkable that the Markov approximation manages to connect the domains for large negative times and large positive times. 

\section{Success and failure of the Markov approximation}
\label{section:5}
The comparison between the Markov solution $a_{M}$ and the exact solution shown in \cref{fig:3} demonstrates in a vivid way that function $\eta_{M}$ reproduces the essential features of $a$. For this reason, we now study $\eta_{M}$ in more detail for different time domains.
\par
In particular, we use elementary properties of integrals to derive a connection formula for $\eta_{M}$, which connects positive and negative times, and brings out most clearly the origin of the Stueckelberg oscillations \cite{Stueckelberg1932}. Moreover, this connection formula allows us to obtain the exact asymptotic value of $a$. However, it fails to predict correctly $b$ due to the lack of the non-linearity.

\subsection{Large negative times}
Throughout this article, we have considered the dynamics of the Landau-Zener problem for initial conditions set at a large but finite negative time, $\tau = -\tau_{0}$, before subsequently taking the limit $\tau_{0} \rightarrow \infty$. Nowhere clearer than for the starting domain of the time evolution, that is, for $-\tau_{0}\lesssim\tau$, do we see that there is a drastic difference between this approach, and one in which we have to set $\tau_{0}=\infty$ from the beginning. In this section, we compare and contrast these two approaches, focusing specifically on their implications for the behavior of $\eta$.
\begin{figure}[htbp]
    \includegraphics[width=\columnwidth]{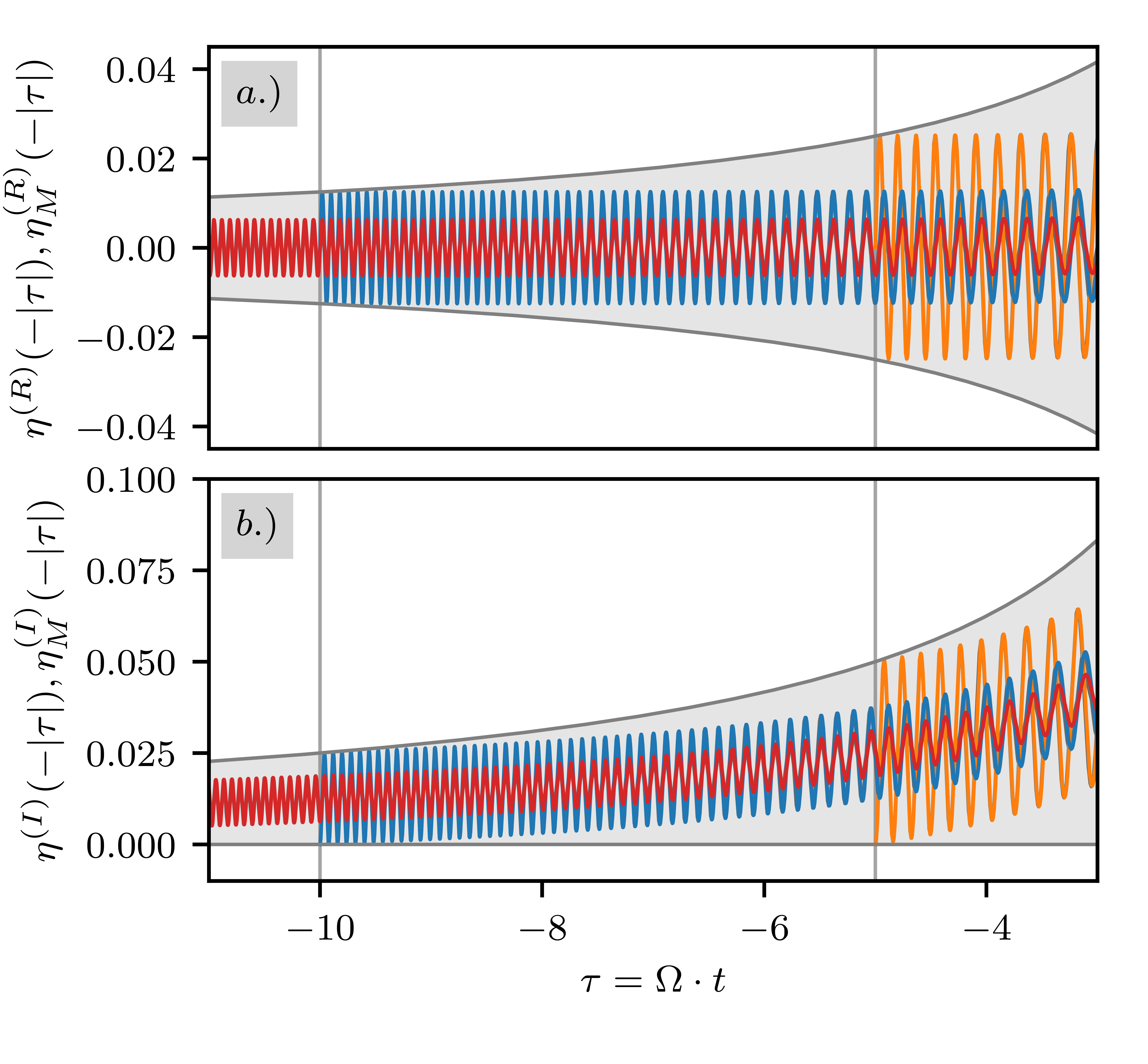}
    \centering
    \caption[]{Oscillations in the functions $\eta$ and $\eta_{M}$, given by \cref{eq:riccati:equation:eta} and \cref{eq:diff:equation:from:eta:M}, for large negative times and for three different starting times $\tau_{0}$ of the Landau-Zener dynamics: (i) $\tau_{0} = 20.0$ (red solid line), (ii) $\tau_{0} = 10.0$ (green solid line), and (iii) $\tau_{0} = 5.0$ (blue solid line). The upper panel $a.)$ illustrates the oscillations in the real parts $\eta^{(R)}$ and $\eta_{M}^{(R)}$, while the lower panel $b.)$ displays the oscillations in the imaginary parts $\eta^{(I)}$ and $\eta_{M}^{(I)}$. The grey solid lines in both panels mark the envelope defined by the initial time $\tau_{0}$, where each point on this envelope corresponds to a potential starting point of the oscillations. The two vertical grey solid lines in panels $a.)$ and $b.)$ indicate the starting points of the oscillations. Here, we have chosen $\epsilon = 4$.}
    \label{fig:9}
\end{figure}

\subsubsection{Emergence of oscillations}
We start our discussion with the case of a vanishing initial condition, \cref{eq:initial:condition:for:eta}, at a finite but large negative time \mbox{$\tau=-\tau_{0}$}. Given that $\eta$ is initially zero, we expect it to remain small during the early-time dynamics. This assumption allows us to linearize the Riccati equation by neglecting its non-linear terms, resulting in the Markov solution $\eta_{M}$ as defined in \cref{eq:definition:markov:function}.
\par
Next, we recall the identity 
\begin{align}
    \int\limits_{-\tau_{0}}^{-\abs{\tau}} \dd\tau^{\prime} \ee^{\ii\epsilon\tau^{\prime 2}} 
    \cong \frac{1}{2\ii\epsilon} \left[ \frac{\ee^{\ii\epsilon\tau_{0}^{2}}}{\tau_{0}} - \frac{\ee^{\ii\epsilon\tau^{2}}}{\abs{\tau}} \right]
    \label{eq:asymptotic:integral:fromula:large:negative:times}
\end{align}
valid for large $\tau_{0}$ as well as in the regime $-\tau_{0}\lesssim\tau$. This result follows directly by substituting $\bar{\tau} \equiv -\tau^{\prime}$ and afterwards $t \equiv \bar{\tau}^{2}$, combined with partial integration.
\par
With the help of the Markov solution, \cref{eq:definition:markov:function}, we arrive at the approximate expression
\begin{align}
    \eta(-\abs{\tau}) \cong \eta_{M}(-\abs{\tau}) \cong \frac{\ii}{2\epsilon}\left[\frac{1}{\abs{\tau}}-\frac{1}{\tau_{0}}\ee^{\ii\epsilon\left(\tau_{0}^{2}-\tau^{2}\right)}\right]
    \label{eq:solution:eta:negative:times:finite}
\end{align}
which satisfies the initial condition, \cref{eq:initial:condition:for:eta}.
\par
Hence, two distinct behaviors govern the early-time dynamics of $\eta_{M}$: 
(i) a growth of $\eta_{M}$ with a dependence $1/\abs{\tau}$ as $\tau\rightarrow 0$, and (ii) an oscillation with a quadratic phase.
\par
Both features manifest themselves in the time dependence of the probability amplitude
\begin{align}
    \begin{split}
        a_{M}(-\abs{\tau})=\ee^{-\ii\epsilon\left(\tau_{0}^{2}-\tau^{2}\right)/2}\ee^{-\mathrm{H}_{M}(-\abs{\tau})}
        \label{eq:a:M:negative:times}
    \end{split}
\end{align}
given by \cref{eq:Markov:solution}.
\par
Indeed, from the the definition, \cref{eq:H:M:definition}, of $\mathrm{H}_{M}$ we find from the approximate expression, \cref{eq:solution:eta:negative:times:finite}, of $\eta_{M}$ the formula
\begin{align}
    \mathrm{H}_{M}(-\abs{\tau}) = \frac{\ii}{2\epsilon}\ln\left(\frac{\abs{\tau}}{\tau_{0}}\right) - \frac{\ii\ee^{\ii\epsilon\tau_{0}^{2}}}{2\epsilon\tau_{0}}\int\limits_{-\tau_{0}}^{-\abs{\tau}}\dd\tau^{\prime}\ee^{-\ii\epsilon\tau^{\prime 2}}
\end{align}
which with the help of the asymptotic formula, \cref{eq:asymptotic:integral:fromula:large:negative:times}, yields 
\begin{align}
    \mathrm{H}_{M}(-\abs{\tau}) \cong \frac{\ii}{2\epsilon}\ln\left(\frac{\abs{\tau}}{\tau_{0}}\right) + \frac{1}{4\epsilon^{2}\tau_{0}^{2}} - \frac{\ee^{\ii\epsilon\left(\tau_{0}^{2} - \tau^{2}\right)}}{4\epsilon^{2}\tau_{0}\abs{\tau}}.  \label{eq:asymptotic:expression:for:H:M:large:negative:times}
\end{align}
\par
Hence, for large negative times we find with \cref{eq:phase:velocity:markov} and \cref{eq:solution:eta:negative:times:finite} the asymptotic phase velocity
\begin{align}
\begin{split}
    \dot{\varphi}_{M}(-\abs{\tau}) &=  -\epsilon\abs{\tau}- \frac{1}{2\epsilon\abs{\tau}}\\
    &+\frac{1}{2\epsilon\tau_{0}}\cos\left[\epsilon\left(\tau_{0}^{2}-\tau^{2}\right)\right].
    \label{eq:precursor:oscillations:phi:dot}
\end{split}
\end{align}
\par
Therefore, the first term in \cref{eq:solution:eta:negative:times:finite} leads to a negative phase velocity which increases as $\tau\rightarrow 0$. In addition, this growth is modulated by an oscillation with an increasing frequency. The amplitude of this oscillation decreases linearly with increasing $\tau_{0}$. Only when $\tau_{0}=\infty$ do these oscillations cease to exist. 
\par
In \cref{fig:9}, we present the dependence of $\eta$ and $\eta_{M}$ on $\tau_{0}$ for three different values of $\tau_{0}$, separately illustrating their real and imaginary parts. The oscillations are clearly visible but their amplitudes decay for increasing $\tau_{0}$ as predicted by \cref{eq:precursor:oscillations:phi:dot}. 
\par
The quadratic phase factor in \cref{eq:solution:eta:negative:times:finite} not only affects the phase velocity but also the amplitude. Indeed, from \cref{eq:definition:A:markov} we find that the amplitude
\begin{align}
    A_{M}(-\abs{\tau}) = \exp\left\{-\frac{1}{2\epsilon\tau_{0}}\int\limits_{-\tau_{0}}^{-\abs{\tau}}\dd\tau^{\prime}\sin\left[\epsilon\left(\tau_{0}^{2}-\tau^{\prime 2}\right)\right]\right\}
\end{align}
also displays oscillations. 
\par
On first sight the appearance of these oscillations is surprising since they manifest themselves already for large negative times, that is long before the coupling between the two levels can take effect. Indeed, we expect the coupling to be most effective in the neighborhood of the level crossing at $\tau=0$ leading to a modulation positive times but not for negative ones. 
\par
\begin{figure}[htbp]
    \centering
    \includegraphics[width=\columnwidth]{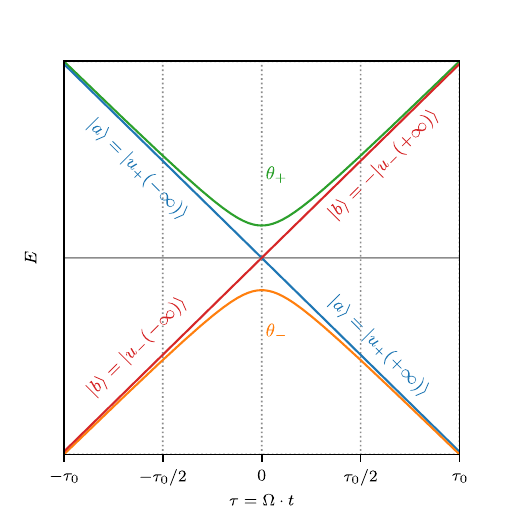}
    \caption{Oscillations in the early time evolution of the probability amplitude $a$ of the Landau-Zener problem resulting from a superposition of instantaneous energy eigenstates. The time-dependent energy levels corresponding to the states, $\ket{a}$ and $\ket{b}$, and displayed over the finite period of time from $-\tau_{0}$ to $\tau_{0}$, decrease (solid blue line) or increase (solid red line) linearly in time with the rate $\epsilon$. They form the diabatic basis. Only in the asymptotic limit, that is $\tau_{0}\rightarrow\infty$, the diabatic and the adiabatic basis represented by the instantaneous eigenvalues, $\theta_{+}$ and $\theta_{-}$,  shown by the green and orange solid lines agree. Hence, an initial condition at a finite time restricted to a single level of the diabatic basis  requires both adiabatic states, $\ket{u_{+}}$ and $\ket{u_{-}}$, and thereby leads to oscillations. Their separation at the origin is a measure of the coupling of the two levels.}
    \label{fig:avoided:crossings}
\end{figure}

However, this picture is misleading since the coupling is always present and creates two instantaneous eigenstates $u_{+}$ and $u_{-}$ of positive and negative eigenvalues $\theta_{+}$ and $\theta_{-}$ as shown in \cref{fig:avoided:crossings}. Only in the limit $\tau\rightarrow\infty$ these eigenvalues approach the levels corresponding to $a$ and $b$ as discussed in more detail in \cref{appendix:B}.
\par
As a result, we can only satisfy the initial condition for $a$ at a finite value of $\tau$ by a superposition of \textit{both} eigenstates giving rise to the oscillations. In particular, we show in \cref{appendix:B} that it is the interference of the two instantaneous eigenstates that leads us to \cref{eq:asymptotic:expression:for:H:M:large:negative:times}. 
\subsubsection{No oscillations}
Deeper insight into the oscillations emerges when we take the first derivative in time of of $\eta_{M}$ given by \cref{eq:solution:eta:negative:times:finite}. The first term, proportional to $1/\tau$ turns into $1/\tau^{2}$ which we neglect. However, due to the quadratic chirp the oscillatory term creates an additional term linear in $\tau$, and as a result, the amplitude is now of the order $\tau/\tau_{0}$, that is of the order $1$. Hence, the existence of the oscillations requires a non-vanishing derivative in time.
\par
In contrast, the solution emerging from the Riccati equation, \cref{eq:riccati:equation:eta}, subjected to the initial condition 
\begin{align}
    \eta(\tau_{0} = -\infty) = 0
    \label{eq:initial:condition:eta:at:infty}
\end{align}
does have a time derivative which solely scales as $1/\tau^{2}$.
\par
In order to demonstrate this fact, we start from the differential equation, \cref{eq:riccati:equation:eta}, in the form
\begin{align}
    \eta(-\tau) = \frac{1}{2\ii\epsilon\tau} + \frac{\eta(\tau)^{2}}{2\ii\epsilon\tau} - \frac{\dot{\eta}(\tau)}{2\ii\epsilon\tau}
    \label{eq:differential:equation:eta:perturbative:version}
\end{align}
and solve it by neglecting the contributions from $\eta^{2}$ and $\dot{\eta}$ which yields
\begin{align}
    \eta(\tau) \cong \frac{1}{2\ii\epsilon\tau}.
    \label{eq:ansatz:eta:large:negative:times}
\end{align}
This approximation satisfies the initial condition, \cref{eq:initial:condition:eta:at:infty} and is free of the oscillations. 
\par 
Indeed, when we differentiate this expression for $\eta$ with respect to time we obtain a contribution proportional to $1/\tau^{2}$ which we neglect. Hence, to this order of the approximation it is justified to neglect the time derivative $\dot{\eta}$ in the Riccati equation, \cref{eq:riccati:equation:eta}. 
Moreover, for $\tau_{0}\rightarrow\infty$ the two solutions, \cref{eq:solution:eta:negative:times:finite} and \cref{eq:ansatz:eta:large:negative:times}, merge.

\subsection{At the time of the transition}
Next, we derive an approximate but analytic expression for $\eta_{M}$ in the neighborhood of $\tau = 0$. For this purpose, we perform a Taylor expansion
\begin{align}
        \eta_{M}(\tau) = \eta_{M}(0) + \dot{\eta}_{M}(0)\tau + \frac{1}{2}\ddot{\eta}_{M}(0)\tau^{2} + \ldots
        \label{eq:taylor:expansion:eta:markov}
\end{align}
of $\eta_{M}$ around $\tau = 0$ and use derivatives of the corresponding differential equation, \cref{eq:riccati:equation:different:representation:markov}, to evaluate the expansion coefficients.
\par
Indeed, from \cref{eq:riccati:equation:different:representation:markov}, we find immediately
\begin{align}
    \dot{\eta}_{M}(0) = 1.  
    \label{eq:eta:M:dot}
\end{align}
\par
When we differentiate \cref{eq:riccati:equation:different:representation:markov} one more time, we arrive at the identity
\begin{align}
     \ddot{\eta}_{M} =- 2\ii\epsilon\eta_{M}- 2\ii\epsilon\tau\dot{\eta}_{M}
\end{align}
which with the help of \cref{eq:eta:M:dot} yields
\begin{align}
    \ddot{\eta}_{M}(0) = -2\ii\epsilon\eta_{M}(0).
\end{align}
\par
\begin{figure}[htbp]
    \includegraphics[width=\columnwidth]{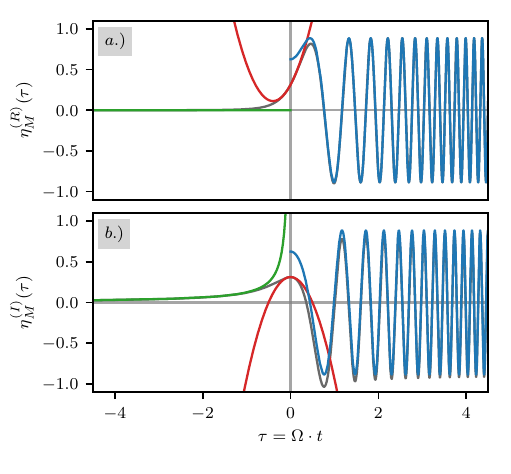}
    \centering
    \caption[]{Comparison between the numerical evaluation of the Markov solution $\eta_{M}$ (grey solid line), and its analytical approximations, \cref{eq:riccati:equation:approximation:large:negative:times}, \cref{eq:taylor:expansion:eta} and \cref{eq:improved:integral:equation:ricatti:markov:1}, in three different time domains:
    (i) Large negative times (green solid line), (ii) at the time of the transition (red solid line), and (iii) large positive times (blue solid line). In $a.)$, we depict the real and in $b.)$ the imaginary parts of the corresponding functions. For our comparison we have chosen the parameter $\epsilon = 4.0$.}
    \label{fig:10}
\end{figure}

Since $\eta_{M}$ is given explicitly we can obtain the expression
\begin{align}
    \eta_{M}(0) = \int\limits_{-\tau_{0}}^{0}\dd\tau\,\ee^{\ii\epsilon\tau^{2}} \equiv \frac{1}{2}\mathcal{F}(\tau_{0}) 
\end{align}
of $\eta_{M}(0)$ where we have introduced the abbreviation
\begin{align}
    \mathcal{F}(\tau_{0}) \equiv \int\limits_{-\tau_{0}}^{\tau_{0}}\dd\tau\,\ee^{\ii\epsilon\tau^{2}}.
    \label{eq:definition:integral:of:F}
\end{align}
The behavior of the Taylor expansion of second order in Markov approximation is depicted in \cref{fig:10} as red solid line.

\subsection{Large positive times}
We now address the behavior of $\eta_{M}$ for large positive times. In order to do so, we first derive a relation that connects the behavior of $\eta_M$ at negative times with its behavior at positive times. This identity is then used to obtain the Stueckelberg oscillations and the exact Landau-Zener result.

\subsubsection{Connection formula}
\label{subsection:1:3:2}
From the definition, \cref{eq:definition:markov:function}, of $\eta_{M}$ we find the expression
\begin{align}
    \eta_{M}(-\tau) = \ee^{-\ii\epsilon\tau^{2}}\int\limits_{-\tau_{0}}^{-\tau}\dd\tau^{\prime}\ee^{\ii\epsilon\tau^{\prime 2}}
\end{align}
which with the new integration variable $\Bar{\tau} \equiv -\tau^{\prime}$ takes the form
\begin{align}
    \eta_{M}(-\tau) = \ee^{-\ii\epsilon\tau^{2}}\int\limits_{\tau}^{\tau_{0}}\dd\Bar{\tau}\,\ee^{\ii\epsilon\Bar{\tau}^{2}}.
\end{align}
\par
Next, we subtract and add an appropriate integral to obtain the relation
\begin{align}
    \begin{split}
            \eta_{M}(-\tau) = \ee^{-\ii\epsilon\tau^{2}}&\left\{-\int\limits_{-\tau_{0}}^{\tau}\dd\Bar{\tau}\,\ee^{\ii\epsilon\Bar{\tau}^{2}} + \int\limits_{\tau}^{\tau_{0}}\dd\Bar{\tau}\,\ee^{\ii\epsilon\Bar{\tau}{2}}\right.\\ &+ \left.\int\limits_{-\tau_{0}}^{\tau}\dd\Bar{\tau}\,\ee^{\ii\epsilon\Bar{\tau}^{2}}\right\}.
    \end{split}
\end{align}
\par
When we combine the last two integrals, we arrive at the identity 
\begin{align}
    \begin{split}
            \eta_{M}(-\tau) = -\eta_{M}(\tau) +\ee^{-\ii\epsilon\tau^{2}}\mathcal{F}(\tau_{0})
    \end{split}
\end{align}
where we have recalled the definitions, \cref{eq:definition:markov:function}, of $\eta_{M}$ and the $\tau$-independent integral, \cref{eq:definition:integral:of:F}, of $\mathcal{F}$.
\par
Hence, we finally arrive at the connection formula
\begin{align}
    \eta_{M}(\tau) = -\eta_{M}(-\tau) + \mathcal{F}(\tau_{0})\ee^{-\ii\epsilon\tau^{2}}
    \label{eq:symmetry:relation:eta:markov:minus}
\end{align}
which demonstrates that $\eta_{M}$ is not anti-symmetric but contains an additional phase factor whose phase increases quadratically with time.

\subsubsection{Stueckelberg oscillations}
Two contributions appear in the connection formula, \cref{eq:symmetry:relation:eta:markov:minus}, of the Markov solution, \cref{eq:Markov:solution}:
(i) For large positive times the term $-\eta_{M}(-\abs{\tau})$ corresponds to $\eta_{M}$ at large negative times, and (ii) phase factor quadratic in time. 
We now show that this phase factor is the origin of the Stueckelberg oscillations. In this derivation, we take advantage of the behavior of $\eta_{M}$ for large negative times. 
\par
For this purpose, we start from the identity
\begin{align}
    \mathrm{H}_{M}(\tau) = \mathrm{H}_{M}(0) + \int\limits_{0}^{\tau}\dd\tau^{\prime}\eta_{M}(\tau^{\prime})
\end{align}
for the integral $\mathrm{H}_{M}$ defined in \cref{eq:H:M:definition} and substitute the connection formula, \cref{eq:symmetry:relation:eta:markov:minus}, into the integral on the right-hand side which yields the relation
\begin{align}
    \begin{split}
        \mathrm{H}_{M}(\tau) = \mathrm{H}_{M}(0)  - \int\limits_{0}^{\tau}\dd\tau^{\prime}\eta_{M}(-\tau^{\prime})+ \mathcal{F}(\tau_{0})\int\limits_{0}^{\tau}\dd\tau^{\prime}\ee^{-\ii\epsilon\tau^{\prime 2}}.
    \end{split}
\end{align}
\par
The substitution $\bar{\tau}=-\tau^{\prime}$ allows us to combine the second integral with $\mathrm{H}_{M}(0)$ and we arrive at the formula
\begin{align}
    \begin{split}
        \mathrm{H}_{M}(\tau) =\mathrm{H}_{M}(-\abs{\tau})+ \mathcal{F}(\tau_{0})\int\limits_{0}^{\tau}\dd\tau^{\prime}\ee^{-\ii\epsilon\tau^{\prime 2}}.
        \label{eq:integral:markov:large:F}
    \end{split}
\end{align}
\par
Next, we recall the Markov solution, \cref{eq:Markov:solution}, 
and use \cref{eq:integral:markov:large:F} which yields the expression
\begin{align}
    \begin{split}
    a_{M}(\abs{\tau}) &= a_{M}(-\abs{\tau})\exp\left(-\mathcal{F}(\tau_{0})\int\limits_{0}^{\tau}\dd\tau^{\prime}\ee^{-\ii\epsilon\tau^{\prime 2}}\right). 
    \label{eq:ansatz:for:diff:eq:a:in:limit:tau:0:rewritten} 
    \end{split}
\end{align}
Here, we have recalled \cref{eq:a:M:negative:times}. 
\par
Hence, due to the connection formula, \cref{eq:symmetry:relation:eta:markov:minus}, of $\eta_{M}$ the behavior of $a_{M}$ for positive times is intimately related to the one at negative times as expressed by the appearance of $a_{M}(-\abs{\tau})$. The origin of this term lies in the contribution $-\eta_{M}(-\abs{\tau})$ in the connection formula. Moreover, the term $\mathcal{F}(\tau_{0})\exp\left(-\ii\epsilon\tau^{2}\right)$ gives rise to the exponential in \cref{eq:ansatz:for:diff:eq:a:in:limit:tau:0:rewritten} which introduces oscillations in the phase as well as the amplitude. These oscillations are called Stueckelberg oscillations and are clearly visible in \cref{fig:6}. 

\subsubsection{Exact Landau-Zener result}
We are now in the position to rederive the exact Landau-Zener result \cite{Zener1932, Landau1932a, Landau1932b, Majorana1932, ivakhnenko2023, Stueckelberg1932, Glasbrenner2023}, \cref{eq:landau:zener:formula:first}. For this purpose, we set in \cref{eq:ansatz:for:diff:eq:a:in:limit:tau:0:rewritten} $\tau = \tau_{0}$ and use the initial condition, \cref{eq:initial:condition:a:markov}, which yields
\begin{align}
    a_{M}(\tau = \tau_{0}) = \exp\left[-\frac{1}{2}\abs{\mathcal{F}(\tau_{0})}^{2}\right].
        \label{eq:short:notation:markov:solution:for:tau:0}
\end{align}
\par
Next, we perform the limit $\tau_{0}\rightarrow\infty$ which with the integral relation
\begin{align}
    \mathcal{F}(\tau_{0}\rightarrow\infty) = \int\limits_{-\infty}^{\infty}\dd\tau\,\ee^{\ii\epsilon\tau^{2}} = \sqrt{\frac{\ii\pi}{\epsilon}},
    \label{eq:integral:relation}
\end{align}
leads us to the expression
\begin{align}
    a_{M}(\tau_{0} \rightarrow \infty) = \exp\left(-\frac{\pi}{2\epsilon}\right) = a(\tau_{0} \rightarrow \infty)
    \label{eq:landau:zener:formula:second}
\end{align}
which is identical to the exact Landau-Zener formula for the probability amplitude $a$.

\subsection{Importance of the non-linearity for \texorpdfstring{$b$}{b}}
The time evolution of the probability amplitudes $a$ and $b$ is governed by the Riccati equation, \cref{eq:riccati:equation:eta}, which is non-linear. In the preceding section, we have demonstrated that this non-linearity is \textit{not} important for the asymptotic value of $a$. However, as we will show now, it \textit{does} play a crucial role in determining the asymptotic value of $b$.
\par
For this purpose, we first derive an expression for $b$ in the Markov approximation which involves the product of the Markov approximations $a_{M}$ and $\eta_{M}$ of $a$ and $\eta$. However, the corresponding expressions yield a result that only in a specific limit agrees with the exact one. We use the non-linear integral relation discussed before to identify the origin of this failure of the Markov approximation.
\par
\subsubsection{Markov approximation for \texorpdfstring{$b$}{b}}
To bring out this fact most clearly, we perform in \cref{eq:formal:solution:for:b} the Markov approximation, \cref{eq:markov:approximation}, which yields
\begin{align}
    b_{M}(\tau) = -\ii\ee^{-\ii\epsilon\tau^{2}}\int\limits_{-\tau_{0}}^{\tau}\dd\tau^{\prime}\ee^{\ii\epsilon\tau^{\prime 2}}a_{M}(\tau),
    \label{eq:formal:solution:for:b:markov:1}
\end{align}
and with the definition, \cref{eq:definition:markov:function}, of $\eta_{M}$, we arrive at the expression
\begin{align}
    b_{M} = -\ii\eta_{M}a_{M}
    \label{eq:expression:for:prob:amplitude:b:a:markov}
\end{align}
for the probability amplitude $b_{M}$ in Markov approximation. 
\par 
We note that we also arrive at this product representation by replacing in \cref{eq:expression:for:prob:amplitude:b:a} all functions by their respective Markov approximations.
\par
Next, we use the product representation, \cref{eq:expression:for:prob:amplitude:b:a:markov}, of $b_{M}$ to calculate $b_{M}(\tau_{0})$. This quantity requires $\eta(\tau_{0})$ which according to \cref{eq:definition:markov:function} is determined by 
\begin{align}
    \eta_{M}(\tau_{0}) =\ee^{-\ii\epsilon\tau_{0}^{2}}\mathcal{F}(\tau_{0}).
\end{align}
Here, we have recalled the definition, \cref{eq:definition:integral:of:F}, of the Fresnel integral.
\par
In the limit $\tau_{0}\rightarrow\infty$, we use the integral relation, \cref{eq:integral:relation}, to find the expression
\begin{align}
    \eta_{M}(\tau_{0}) = \ee^{-\ii\epsilon\tau_{0}^{2}}\ee^{\ii\pi/4}\sqrt{\frac{\pi}{\epsilon}}.
\end{align}
\par
When we substitute this result together with the formula \cref{eq:landau:zener:formula:second} for $a_{M}(\tau_{0})$ into the product representation, \cref{eq:expression:for:prob:amplitude:b:a:markov}, of $b_{M}$, we find
\begin{align}
    b_{M}(\tau = \tau_{0} \rightarrow\infty) = \ee^{-\pi/(2\epsilon)}\sqrt{\frac{\pi}{\epsilon}}\ee^{-\ii\left(\epsilon\tau_{0}^{2} + \pi/4\right)}
\end{align}
which in general does not agree with the result, \cref{eq:b:literature}, obtained from the asymptotics of the parabolic cylinder functions.
\par
In particular, the absolute value
\begin{align}
    \abs{b_{M}(\tau = \tau_{0} \rightarrow\infty)} = \sqrt{\frac{\pi}{\epsilon}}\ee^{-\pi/(2\epsilon)}
\end{align}
does not agree with the exact result
\begin{align}
    \begin{split}
        \abs{b(\tau= \tau_{0} \rightarrow\infty)} = \left(1 - \ee^{-\pi/\epsilon}\right)^{\frac{1}{2}}.
    \end{split}
\end{align}
given by \cref{eq:b:literature}. 
\par
Only in the limit of $1\ll\epsilon$ do we find by expanding the exponential 
\begin{align}
    \ee^{-\pi/\epsilon} \cong 1 - \frac{\pi}{\epsilon}
\end{align}
the identity
\begin{align}
    \abs{b(\tau = \tau_{0} \rightarrow\infty)} \cong \abs{b_{M}(\tau = \tau_{0} \rightarrow\infty)}.
\end{align}
\par
Moreover, the phase of $b_{M}$ differs significantly from the exact one given by \cref{eq:phi:eta:bar:infty}, even though the contribution quadratic in $\tau_{0}$ appears in both expressions. Indeed, the logarithmic term as well as the argument of the $\Gamma$-function are missing. Additionally, the Markov approximation predicts a phase shift of $-\pi/4$ rather than $3\pi/4$.

\subsubsection{No factorization of the Markov approximation}
The origin of the failure of the Markov approximation to predict correctly the asymptotic value of $b$ stands out most clearly when we express in \cref{eq:formal:solution:for:b} the probability amplitude $a$ by the exponential ansatz, \cref{eq:ansatz:for:diff:eq:a}, which yields 
\begin{align}
    \begin{split}
        &b(\tau) = -\ii\ee^{-\ii\epsilon\tau_{0}^{2}/2}\\&\times\ee^{-\ii\epsilon\tau^{2}/2}\int\limits_{-\tau_{0}}^{\tau}\dd\tau^{\prime}\ee^{\ii\epsilon\tau^{\prime 2}}\exp\left(-\int\limits_{-\tau_{0}}^{\tau^{\prime}}\dd\tau^{\prime\prime}\eta(\tau^{\prime\prime})\right).
        \label{eq:b:with:exp:ansatz:for:a}
    \end{split}
\end{align}
\par
The Markov approximation replaces the upper limit of the integral over $\tau^{\prime\prime}$ by $\tau$, replacing the integral over $\tau^{\prime}$ by the product of two integrals, that is
\begin{align}
    \begin{split}
        &b_{M}(\tau) = -\ii\ee^{-\ii\epsilon\tau^{2}}\int\limits_{-\tau_{0}}^{\tau}\dd\tau^{\prime}\ee^{\ii\epsilon\tau^{\prime 2}}\\&\times \ee^{-\ii\epsilon(\tau_{0}^{2}-\tau^{2})/2}\exp\left(-\int\limits_{-\tau_{0}}^{\tau}\dd\tau^{\prime\prime}\eta_{M}(\tau^{\prime\prime})\right)
        \label{eq:b:with:exp:ansatz:for:a:markov}
    \end{split}
\end{align}
which with the definitions, \cref{eq:definition:markov:function} and \cref{eq:ansatz:for:diff:eq:a}, of $\eta_{M}$ and $a_M$ yields obviously \cref{eq:expression:for:prob:amplitude:b:a:markov}.
\par
At the very heart of this failure of the Markov approximation to correctly capture the asymptotic expression for $b$ lies the well-known fact that the integral of the product of two complex-valued functions is not necessarily equal to the product of the integrals of each function, multiplied by the other function. In particular, the phases of the two expressions may differ significantly, as demonstrated in this example. However, in the limit $1\ll\epsilon$, the Markov approximation recovers the correct result.

\section{Riccati equation: Asymptotics}
\label{section:6}
In \cref{section:3}, we have obtained the Riccati equation for the integrand $\eta$ of our exponential ansatz, \cref{eq:ansatz:for:diff:eq:a}. In the present section, we now derive approximate analytic results for $\eta$ in three domains of time, and compare and contrast them to the corresponding expressions for $\eta_{M}$. This analysis brings out most clearly the influence of the non-linearity of the Riccati equation compared to the linearized Markov solution.
\par
Throughout this section, we focus on three time domains each of which requires a different approximation:
(i) An elementary iteration of the Riccati equation provides us immediately with an approximate expression of $\eta$ for large \textit{negative} times. (ii) A Taylor expansion, taking advantage of the Riccati equation, \cref{eq:riccati:equation:eta} provides us with an expression around $\tau = 0$, and (iii) the connection formula, \cref{eq:symmetry:relation:eta:markov:minus}, allows us to obtain approximate but analytic results for large \textit{positive} times taking advantages of the knowledge of $\eta$ for large negative times.

\begin{figure}[ht]
    \centering
    \includegraphics[width=\the\columnwidth]{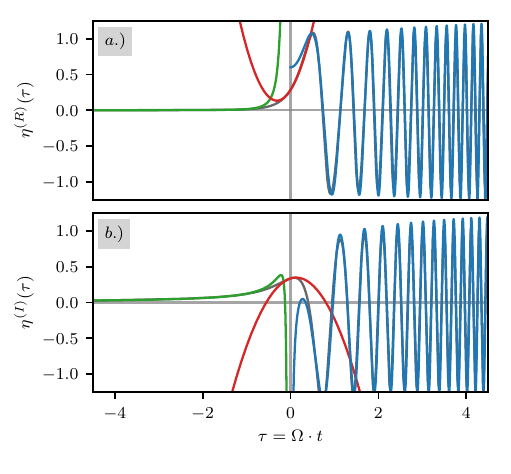}
    \caption{Comparison between the numerical solution $\eta$ (grey solid line) of the Riccati equation, and its analytical approximations, \cref{eq:riccati:equation:approximation:large:negative:times}, \cref{eq:taylor:expansion:eta} and \cref{eq:improved:integral:equation:ricatti:markov:1}, in three different time domains:
    (i) Large negative times (green solid line), (ii) at the time of the transition (red solid line), and (iii) large positive times (blue solid line). In $a.)$, we depict the real and in $b.)$ the imaginary parts of the corresponding functions. For our comparison we have chosen the parameter $\epsilon = 4.0$.}
    \label{fig:enter-label}
\end{figure}

\subsection{Large negative times}
\label{section:large:negative:times}
In the preceding section, we have concentrated on the lowest approximation of the Riccati equation, \cref{eq:riccati:equation:eta}, given by the Markov solution, \cref{eq:Markov:solution}. We now discuss the influence of the non-linearity contained in the quadratic term of the Riccati equation, \cref{eq:riccati:equation:eta}. For the sake of simplicity we demonstrate this feature for the initial condition, \cref{eq:initial:condition:eta:at:infty}.
\par
For this purpose, we insert this ansatz, \cref{eq:ansatz:eta:large:negative:times}, into the right-hand side of \cref{eq:differential:equation:eta:perturbative:version} leading us to the approximation
\begin{align}
    \eta_{R}(\tau) = \frac{1}{2\ii\epsilon\tau}-\frac{1}{4\epsilon^{2}\tau^{3}}-\frac{1}{8\ii\epsilon^{3}\tau^{3}}.
    \label{eq:riccati:equation:approximation:large:negative:times}
\end{align}
Here, the subscript $R$ indicates that we have obtained this expression by an iteration of the Riccati differential equation \cref{eq:riccati:equation:eta}.

We emphasize that the cubic correction to the imaginary part
\begin{align}
    \eta^{(I)}_{R}(\tau) =  -\frac{1}{2\epsilon\tau} + \frac{1}{8\epsilon^{3}\tau^{3}}
\end{align}
is a consequence of the non-linearity, that is 
\begin{align}
    \eta^{(I)}_{R}(\tau) = \eta_{M, D}^{(I)}(\tau) + \frac{1}{8\epsilon^{3}\tau^{3}}.
\end{align}
\par
Hence, for large negative times neglecting the cubic terms the phase velocity calculated from $\eta$ and $\eta_{M}$ agree, \cref{eq:riccati:equation:eta} and \cref{eq:diff:equation:from:eta:M}.
\par
However, close to $\tau = 0$ the cubic correction to $\eta_{M}$ becomes important and marks the onset of the transition.
\par
In contrast to the phase velocity, the absolute value $\abs{a}$ given by the real part, that is 
\begin{align}
    \eta^{(R)}_{R}(\tau) = -\frac{1}{4\epsilon^{2}\tau^{3}} = \eta_{M,R}^{(R)}
\end{align}
is not influenced by the non-linearity in this order of the approximation. 
\par
The behaviour of the real and imaginary part of $\eta_{R}$ are depicted in \cref{fig:enter-label} in the panels $a.)$ and $b.)$ as green solid line. 

\subsection{At the time of the transition}
The previous analysis in \cref{section:large:negative:times} has already indicated that the non-linearity becomes important in the neighborhood of $\tau = 0$, that is at the time of the transition. In order to bring this feature to light we now perform a Taylor expansion
\begin{align}
        \eta(\tau) = \eta(0) + \dot{\eta}(0)\tau + \frac{1}{2}\ddot{\eta}(0)\tau^{2} + \ldots
        \label{eq:taylor:expansion:eta}
\end{align}
of $\eta$ around $\tau = 0$ and use derivatives of the Riccati equation, \cref{{eq:riccati:equation:eta}}, to evaluate the expansion coefficients. We compare and contrast these coefficients to the corresponding ones of $\eta_{M}$.
\par
From the Riccati equation, \cref{eq:riccati:equation:eta}, we find immediately 
\begin{align}
    \dot{\eta}(0) = 1 + \eta(0)^{2} 
    \label{eq:eta:taylor:expansion:lowest:order}
\end{align}
which for $\eta_{M}$ reduces to
\begin{align}
    \dot{\eta}_{M}(0) = 1.  
\end{align}
\par
When we differentiate the Riccati equation one more time we arrive at the identity
\begin{align}
    2\eta\dot{\eta} - \ddot{\eta} - 2\ii\epsilon\eta- 2\ii\epsilon\tau\dot{\eta} = 0
\end{align}
which with the help of \cref{eq:eta:taylor:expansion:lowest:order} yields
\begin{align}
    \ddot{\eta}(0) = 2\eta(0)\left[1 + \eta(0)^{2} - \ii\epsilon\right].
\end{align}
In contrast, the corresponding expression for $\eta_{M}$ reads
\begin{align}
    \ddot{\eta}_{M}(0) = -2\ii\epsilon\eta_{M}(0).
\end{align}
This elementary analysis shows clearly the important role of the non-linearity in the neighborhood of $\tau=0$.
\par
Since $\eta_{M}$ is given explicitly we can obtain the expression
\begin{align}
    \eta_{M}(0) = \int_{-\infty}^{0}\dd\tau\ee^{\ii\epsilon\tau^{2}} = \frac{1}{2}\sqrt{\frac{\ii\pi}{\epsilon}} 
\end{align}
of $\eta_{M}(0)$. In contrast, no such expression exists for $\eta(0)$. 
\par
The behavior of the Taylor expansion of second order in Markov approximation is depicted in \cref{fig:enter-label} as the red solid line.

\subsection{Large positive times}
In \cref{fig:3}, we have found a qualitative agreement between the functions $\eta_{M}$ and $\eta$. However, there is a slight shift in the phase, and the amplitudes of the oscillations for positive times are smaller in $\eta_{M}$ than in $\eta$.

This modification of $\eta_{M}$ stands out most clearly in the integral form, \cref{eq:integral:equation:riccati:equation}, of the Riccati equation. Indeed, $\eta_{M}$ is only the lowest approximation neglecting the non-linearity. 

Hence, an improved solution follows by replacing $\eta^{2}$ in the integral by $\eta_{M}^{2}$, that is 
\begin{align}
    \eta_{I}(\tau) = \eta_{M}(\tau) + \ee^{-\ii\epsilon\tau^{2}}\int\limits_{-\tau_{0}}^{\tau}\dd\tau^{\prime}\eta_{M}(\tau^{\prime})^{2}\ee^{\ii\epsilon\tau^{\prime 2}}
    \label{eq:improved:integral:equation:ricatti:markov}
\end{align}
where the subscript $I$ indicates that we have now iterated the integral equation \cref{eq:integral:equation:riccati:equation}, of the Riccati equation to obtain an expression valid for large positive times. 
\par
We gain more insight into this expression when we recall the connection formula, \cref{eq:symmetry:relation:eta:markov:minus}, which casts 
\cref{eq:improved:integral:equation:ricatti:markov} into the form 
\begin{align}
    \begin{split}
        \eta_{I}(\tau) &= -\eta_{M}(-\abs{\tau})\\ &+ \ee^{-\ii\epsilon\tau^{2}}\left[\mathcal{F}(\tau_{0})
        + \int\limits_{-\tau_{0}}^{\tau}\dd\tau^{\prime}\eta_{M}(\tau^{\prime})^{2}\ee^{\ii\epsilon\tau^{\prime 2}}\right].
    \end{split}
    \label{eq:improved:integral:equation:ricatti:markov:1}
\end{align}
Since according to \cref{eq:ansatz:eta:large:negative:times} $\eta_{M}$ is of the form,
\begin{align}
    \eta_{M}(-\abs{\tau}) = -\frac{1}{2\ii\epsilon\abs{\tau}}
\end{align}
that it decays for increasing time we find the approximation
\begin{align}
    \begin{split}
        \eta_{I}(\tau) &\cong \ee^{-\ii\epsilon\tau^{2}}\left[\mathcal{F}(\tau_{0})
        + \int\limits_{-\tau_{0}}^{\tau}\dd\tau^{\prime}\eta_{M}(\tau^{\prime})^{2}\ee^{\ii\epsilon\tau^{\prime 2}}\right].
    \end{split}
    \label{eq:improved:integral:equation:ricatti:markov:1:large:times}
\end{align}
Hence, the strength of the oscillatory term which in $\eta_{M}$ is solely given by $\mathcal{F}(\tau_{0})$ in the brackets, is modified by the integral containing $\eta_{M}^2$ and a quadratic phase factor. Since this is a complex number with an amplitude and phase, we find a modification of the amplitude as well as the phase.
\par
The behavior of $\eta_{I}(\tau)$ is portrayed in \cref{fig:enter-label} as a blue solid line.
 
\section{Conclusion and outlook}
\label{section:7}

The two probability amplitudes $a$ and $b$ in the well-known Landau-Zener problem follow from the complex-valued solution $\eta$ of a single non-linear differential equation of first order in the form of a Riccati equation. The function $\eta$ is given either by its real and imaginary part $\eta^{(R)}$ and $\eta^{(I)}$, or its absolute value $\abs{\eta}$ and its phase $\phi_{\eta}$.
\par
The way in which these quantities enter into the expressions for $a$ and $b$ is rather intriguing. Indeed, the absolute value $\abs{a}$ is determined by the time integral of $\eta^{(R)}$ which due to the normalization condition also defines $\abs{b}$. Both amplitudes share a common phase $\varphi$ determined by the difference of a phase quadratic in time and the time integral of $\eta^{(I)}$. However, $b$ enjoys an additional phase factor given by $\phi_{\eta}$.
\par
It is interesting to note that $\abs{\eta}$ does not appear in the expressions for the probability amplitudes $a$ and $b$. This feature is a consequence of the fact that $b$ is proportional to the product of $a$ and $\eta$, combined with the normalization condition.
\par
Hence, there exists an asymmetry between $a$ and $b$. Whereas their absolute values are determined by the time integral of $\eta^{(R)}$, the phase difference between $a$ and $b$ does not involve a time integral of $\eta$ but the phase $\phi_{\eta}$, governed by the ratio $\eta^{(I)}/\eta^{(R)}$. 
\par
The Riccati equation of the Landau-Zener problem has also provided us with a fresh point of view on the Markov approximation pursued in Ref. \cite{Glasbrenner2023, Glasbrenner2024book}. Indeed, factoring the probability amplitude $a$ out of the integral is equivalent to neglecting the quadratic contribution in the Riccati equation. Several approximate but analytic expressions for the solution of the Riccati equation in the different time domains have allowed us to gain deeper insight into the validity of the Markov approximation and into the dynamics of a Landau-Zener transition.
\par
Nevertheless, many questions remain. Three examples may suffice to illustrate this point.
\par
Our analysis suggests that the non-linearity in the Riccati equation does not contribute to the long-time limit of the transition probability $\abs{a}^2$. What is the reason for this unusual behavior?
\par
In contrast, the expression for $b$, when we neglect the non-linearity, is drastically different from the well-known formula. Indeed, we have identified the non-Markovian nature of the dynamics of $b$, contained in the non-linear integral equations for $\eta$, as the origin of this failure of the Markov approximation. Is it possible to obtain the familiar asymptotic phase $\phi$ from this integral equation?
\par
The standard approach towards the Landau-Zener problem relies on the linear differential equation of second order leading to the parabolic functions as solutions. This equation corresponds to the Newton equation of a harmonic oscillator \cite{Briggs2013} with a linear time dependence of the frequency and a complex-valued time-independent offset. As a result, two real-valued harmonic oscillators with time-dependent frequencies, coupled in a non-reciprocal way, emerge. Is it possible to rederive the Landau-Zener formula from this picture?
\par
Although, we have gained numerous answers to and many insights into some aspects of these questions, we are confident that there are many more interesting discoveries to be made in this paradigm of quantum mechanics.

\begin{acknowledgments}
We thank G. S. Agarwal, P. Boegel, M. Efremov, D. Fabian, A. Friedrich, V. V. Kocharovsky, T. Reisser, M. O. Scully, J. Seiler, M. E. N. Tschaffon and W. G. Unruh for many fruitful discussions.
W.P.S. is most grateful to Texas A\&M University for a Faculty Fellowship at the Hagler Institute for Advanced Study at Texas A\& M University and to Texas A\&M AgriLife for the support of this work. The research of the IQST is financially supported by the Baden-Württemberg Ministry of Science, Research and Arts.
\end{acknowledgments}

\appendix

\section{Approach based on parabolic cylinder functions}
\label{appendix:A}
In this appendix, we rederive for the sake of completeness, expressions for the complex-valued probability amplitudes $a$ and $b$, valid for arbitrary times using parabolic cylinder functions \cite{Zener1932}. 
Moreover, we obtain an analogous expression for the solution $\eta$ of the Riccati equation presented in \cref{section:3}. These formulae allow us to obtain the familiar results, \cref{eq:landau:zener:formula:first} and \cref{eq:b:literature}, of the asymptotic limit. Here, we pay special attention to the derivation of the phase $\phi$ given by \cref{eq:phi:eta:bar:infty}.
\par
Throughout this appendix we follow the analysis of the main body of this article, that is we first derive the time-dependent solution of $a$, and then obtain the corresponding expression for $b$ by differentiation. From the ratio of $a$ and $b$ we find $\eta$. Although, our expressions are valid for arbitrary initial conditions, we also address the special ones discussed in the article. Finally, the familiar asymptotic expansions \cite{abramowitz1965, gradshteyn2014} of the parabolic cylinder functions reduce these expressions to the well-known Landau-Zener results. 
\par
We feel justified revisiting this celebrated calculation since in the literature not only various scalings  but also different initial conditions have been used which makes it difficult to compare expressions. Moreover, many subtleties of this derivation have been hidden behind an opaque curtain of mathematics. 
\par
\subsection{Expression for probability amplitude \texorpdfstring{$a$}{a}}
We start our discussion by recalling the linear second order differential equation
\begin{align}
    \label{eq:standard:form:diff:eq:a:app}
    \dv[2]{z}a(z) - \left[\frac{z^{2}}{4} + \frac{\ii}{2\epsilon} + \frac{1}{2} \right]a(z) = 0
\end{align}
for the probability amplitudes $a$ with the complex scaling
\begin{align}
     z \equiv \sqrt{2\epsilon}\tau\ee^{\ii\frac{\pi}{4}}.
     \label{eq:complex:scaling:appendix}
\end{align}
We emphasize, that $z$ differs from the corresponding quantity in Ref. \cite{ivakhnenko2023} by the factor $\sqrt{\epsilon}$. 
\par
According to Ref. \cite{abramowitz1965, gradshteyn2014} the combination 
\begin{align}
    a(z) = \mathcal{A}_{+}\mathcal{D}_{\nu}(z)+\mathcal{A}_{-}\mathcal{D}_{\nu}(-z)
    \label{eq:ansatz:appendix:a}
\end{align}
of parabolic cylinder functions $\mathcal{D}_{\nu}$ with arguments $z$ and $-z$ and order
\begin{align}
    \nu \equiv -1-\frac{\ii}{2\epsilon}
    \label{eq:definition:of:nu:appendix}
\end{align}
is a solution to \cref{eq:standard:form:diff:eq:a:app}. Here, $\mathcal{A}_{\pm}$ are constants to match the initial conditions \cref{eq:initial:condition:a} and \cref{eq:initial:condition:a:dot} which in the complex scaling \cref{eq:complex:scaling:appendix} read
\begin{align}
    a(-z_{0}) = 1
    \label{eq:initial:condition:appendix:a:z}
\end{align}
and 
\begin{align}
    \dv{z}a(-z_{0}) = -\frac{z_0}{2}.
    \label{eq:initial:condition:appendix:a:dot:z}
\end{align}
\par
We first derive expressions for $\mathcal{A}_{\pm}$ for arbitrary initial conditions and then reduce them to the ones given by \cref{eq:initial:condition:appendix:a:z} and \cref{eq:initial:condition:appendix:a:dot:z}.
\par
From the ansatz, \cref{eq:ansatz:appendix:a}, for $a$ we find the relation 
\begin{align}
    a(-z_{0}) = \mathcal{A}_{+}\mathcal{D}_{\nu}(-z_{0})+\mathcal{A}_{-}\mathcal{D}_{\nu}(z_{0}),
    \label{eq:ansatz:appendix:a:with:initial:values}
\end{align}
and differentiation of \cref{eq:ansatz:appendix:a} leads us to the expression
\begin{align}
    \dv{z}a(z) = \mathcal{A}_{+}\parabolicfunctiondot{\nu}{z}-\mathcal{A}_{-}\parabolicfunctiondot{\nu}{-z}
    \label{eq:ansatz:appendix:a:dot}
\end{align}
where 
\begin{align}
    \parabolicfunctiondot{\nu}{z} \equiv \dv{\parabolicfunction{\nu}{\xi}}{\xi}\eval_{\xi=z}.
\end{align}
We note that the second term in \cref{eq:ansatz:appendix:a:dot} results from the chain rule 
\begin{align}
    \dv{z}\parabolicfunction{\nu}{-z} = (-1)\cdot\dv{\parabolicfunction{\nu}{\xi}}{\xi}\eval_{\xi=-z} \equiv -\parabolicfunctiondot{\nu}{-z}.
\end{align}
\par
As a consequence, we find 
\begin{align}
    \dv{z}a(-z_{0}) = \mathcal{A}_{+}\parabolicfunctiondot{\nu}{-z_{0}}-\mathcal{A}_{-}\parabolicfunctiondot{\nu}{z_{0}}
    \label{eq:ansatz:appendix:a:prime}
\end{align}
which together with \cref{eq:ansatz:appendix:a:with:initial:values} can be cast into the vector form
\begin{align}
    \begin{pmatrix}
        a(-z_{0}) \\
        \dv{z}a(-z_{0})
    \end{pmatrix} = M
    \begin{pmatrix}
        \mathcal{A}_{+} \\
        \mathcal{A}_{-}
    \end{pmatrix} 
    \label{eq:matrix:equation}
\end{align}
where
\begin{align}
    M \equiv
    \begin{pmatrix}
       \parabolicfunction{\nu}{-z_{0}} & \parabolicfunction{\nu}{z_{0}}\\
       \parabolicfunctiondot{\nu}{-z_{0}} & -\parabolicfunctiondot{\nu}{z_{0}}
    \end{pmatrix}.
\end{align}
\par
Next, we solve \cref{eq:matrix:equation} for the constants $\mathcal{A}_{\pm}$ by inverting $M$ which yields the identity
\begin{align}
    \begin{pmatrix}
        \mathcal{A}_{+} \\
        \mathcal{A}_{-}
    \end{pmatrix} = M^{-1}
    \begin{pmatrix}
        a(-z_{0}) \\
        \dv{z}a(-z_{0})
    \end{pmatrix} 
\end{align}
where 
\begin{align}
    M^{-1} = \frac{1}{\mathcal{M}}   
    \begin{pmatrix}
       -\parabolicfunctiondot{\nu}{z_{0}} & -\parabolicfunction{\nu}{z_{0}}\\
       -\parabolicfunctiondot{\nu}{-z_{0}} & \parabolicfunction{\nu}{-z_{0}}
    \end{pmatrix}.
\end{align}
denotes the inverse matrix of $M$ and 
\begin{align}
    \begin{split}
        \mathcal{M} &\equiv \det\left(M\right)\\
        &= -\left[\parabolicfunction{\nu}{-z_{0}}\parabolicfunctiondot{\nu}{z_{0}}+\parabolicfunction{\nu}{z_{0}}\parabolicfunctiondot{\nu}{-z_{0}}\right]
    \end{split}
\end{align}
is its determinant.
\par
As a result, we find the constants
\begin{align}
    \mathcal{A}_{+} = -\frac{1}{\mathcal{M}}\left[\parabolicfunctiondot{\nu}{z_{0}}a(-z_{0})+\parabolicfunction{\nu}{z_{0}}\dv{z}a(-z_{0})\right]
\end{align}
and
\begin{align}
    \begin{split}
        \mathcal{A}_{-} = -\frac{1}{\mathcal{M}}&\left[\parabolicfunctiondot{\nu}{-z_{0}}a(-z_{0})\right.\\&-\left.\parabolicfunction{\nu}{-z_{0}}\dv{z}a(-z_{0})\right].
    \end{split}
\end{align}
\par
When we replace the derivatives of the parabolic cylinder function using the relation \cite{gradshteyn2014}
\begin{align}
    \parabolicfunctiondot{\nu}{z}= \frac{z}{2}\parabolicfunction{\nu}{z} - \parabolicfunction{\nu + 1}{z}.
    \label{eq:relation:parabolic:cylinder:function:plus:z}
\end{align}
the determinant $\mathcal{M}$ and the constants $\mathcal{A}_{\pm}$ take the form
\begin{align}
    \mathcal{M} = \parabolicfunction{\nu}{-z_{0}}\parabolicfunction{\nu+1}{z_{0}}+\parabolicfunction{\nu}{z_{0}}\parabolicfunction{\nu+1}{-z_{0}}.
    \label{eq:expression:for:M}
\end{align}
and
\begin{align}
    \begin{split}
        \mathcal{A}_{+} = \frac{1}{\mathcal{M}}&\left\{\parabolicfunction{\nu+1}{z_{0}}a(-z_{0})\right.\\
        &\left.-\parabolicfunction{\nu}{z_{0}}\left[\frac{z_{0}}{2}+\dv{z}a(-z_{0})\right]\right\}
        \label{eq:a:plus:constant:1}
    \end{split}
\end{align}
together with
\begin{align}
    \begin{split}
        \mathcal{A}_{-} = \frac{1}{\mathcal{M}}&\left\{\parabolicfunction{\nu+1}{-z_{0}}a(-z_{0})\right.\\&+\left.\parabolicfunction{\nu}{-z_{0}}\left[\frac{z_{0}}{2}+\dv{z}a(-z_{0})\right]\right\}.
        \label{eq:a:minus:constant:1}
    \end{split}
\end{align}
\par
So far, we have found an expression for the probability amplitude $a$ for arbitrary initial conditions. We now specify the solution for the initial conditions, \cref{eq:initial:condition:appendix:a:z} and \cref{eq:initial:condition:appendix:a:dot:z}, and \cref{eq:a:plus:constant:1} and \cref{eq:a:minus:constant:1} reduce to 
\begin{align}
    \label{eq:A:plus:small}
    \mathcal{A}_{+} &= \frac{\parabolicfunction{\nu+1}{z_{0}}}{\mathcal{M}}\\
    \mathcal{A}_{-} &= \frac{\parabolicfunction{\nu+1}{-z_{0}}}{\mathcal{M}}.
    \label{eq:A:minus:small}
\end{align}
Thus, we obtain the expression
\begin{align}
    \begin{split}
         a(z) = \frac{\parabolicfunction{\nu+1}{z_{0}}\parabolicfunction{\nu}{z}+\parabolicfunction{\nu+1}{-z_0}\parabolicfunction{\nu}{-z}}{\parabolicfunction{\nu}{-z_{0}}\parabolicfunction{\nu+1}{z_{0}}+\parabolicfunction{\nu}{z_{0}}\parabolicfunction{\nu+1}{-z_{0}}}
         \label{eq:a:of:z:parabolic:cylinder:function}
    \end{split}
\end{align}
for the probability amplitude $a$. Here, we have recalled \cref{eq:expression:for:M} for $\mathcal{M}$.

\subsection{Expression for probability amplitude \texorpdfstring{$b$}{b}}
In order to derive an expression for $b$ we recall \cref{eq:diff:equation:a} which in the complex scaling, \cref{eq:complex:scaling:appendix}, reads
\begin{align}
    \ii\frac{\dd}{\dd z}a(z) = \ii\frac{z}{2}a(z) - \frac{\ee^{-\ii\frac{\pi}{4}}}{\sqrt{2\epsilon}}b(z),
    \label{eq:differential:equation:first:order:for:a}
\end{align}
and cast it into the form
\begin{align}
    b(z) = \sqrt{2\epsilon}\ee^{-\ii\frac{\pi}{4}}\left[\dv{z}a(z) -\frac{z}{2}a(z)\right].
\end{align}
\par
Together with the expressions, \cref{eq:ansatz:appendix:a} and \cref{eq:ansatz:appendix:a:dot}, for $a$ and its derivative, we find
\begin{align}
    \begin{split}
        b(z) = \sqrt{2\epsilon}\ee^{-\ii\frac{\pi}{4}}&\left[\mathcal{A}_{+}\left(\parabolicfunctiondot{\nu}{z}-\frac{z}{2}\parabolicfunction{\nu}{z}\right)\right.\\&\left.-\mathcal{A}_{-}\left(\parabolicfunctiondot{\nu}{-z}+\frac{z}{2}\parabolicfunction{\nu}{-z}\right)\right]
    \end{split}
\end{align}
which with the help of the relation, \cref{eq:relation:parabolic:cylinder:function:plus:z}, for the derivative of the parabolic cylinder function, reduces to
\begin{align}
    \begin{split}
        b(z) = -\sqrt{2\epsilon}\ee^{-\ii\frac{\pi}{4}}\left[\mathcal{A}_{+}\parabolicfunction{\nu+1}{z}-\mathcal{A}_{-}\parabolicfunction{\nu+1}{-z}\right].
    \end{split}
\end{align}
\par
We now return to the initial conditions, \cref{eq:initial:condition:appendix:a:z} and \cref{eq:initial:condition:appendix:a:dot:z}, for which the constants $\mathcal{A}_{\pm}$ are given by \cref{eq:A:plus:small} and \cref{eq:A:minus:small}. Thus, we arrive at the expression 
\begin{align}
    \begin{split}
        b(z) &= -\sqrt{2\epsilon}\ee^{-\ii\frac{\pi}{4}}\\ &\times\frac{\parabolicfunction{\nu+1}{z_{0}}\parabolicfunction{\nu+1}{z}-\parabolicfunction{\nu+1}{-z_{0}}\parabolicfunction{\nu+1}{-z}}{\parabolicfunction{\nu}{-z_{0}}\parabolicfunction{\nu+1}{z_{0}}+\parabolicfunction{\nu}{z_{0}}\parabolicfunction{\nu+1}{-z_{0}}}
        \label{eq:b:of:z:parabolic:cylinder:function}
    \end{split}
\end{align}
for the probability amplitude $b$ which satisfies the initial condition $b(-z_{0})=0$, in complete agreement with \cref{eq:initial:condition:b}.

\subsection{Expression for solution \texorpdfstring{$\eta$}{eta} of the Riccati equation}
In \cref{section:3}, we have presented an ansatz, \cref{eq:ansatz:for:diff:eq:a}, for the probability amplitude $a$ introducing the function $\eta$ which leads us directly to the non-linear differential equation of first order, \cref{eq:riccati:equation:eta}, for $\eta$ of the Riccati type. We now derive a representation of $\eta$ in terms of the parabolic cylinder functions.
\par
For this purpose, we first recall \cref{eq:expression:for:prob:amplitude:b:a} which in the complex scaling provides us with the expression
\begin{align}
    \eta(z) = \ii\frac{b(z)}{a(z)}
    \label{eq:appendix:eta:b:over:a}
\end{align}
for the function $\eta$.
\par
Next, we insert the representations, \cref{eq:a:of:z:parabolic:cylinder:function} and \cref{eq:b:of:z:parabolic:cylinder:function} of the probability amplitudes $a$ and $b$ into \cref{eq:appendix:eta:b:over:a} and arrive at
\begin{align}
    \begin{split}
        \eta(z) &= -\sqrt{2\epsilon}\ee^{-\ii\frac{\pi}{4}}\\ &\times\frac{\parabolicfunction{\nu+1}{z_{0}}\parabolicfunction{\nu+1}{z}-\parabolicfunction{\nu+1}{-z_{0}}\parabolicfunction{\nu+1}{-z}}{\parabolicfunction{\nu+1}{z_{0}}\parabolicfunction{\nu}{z}+\parabolicfunction{\nu+1}{-z_{0}}\parabolicfunction{\nu}{-z}}
        \label{eq:representation:eta:parabolic:cylinder:functions}
    \end{split}
\end{align}
which satisfies the initial condition $\eta(-z_{0})=0$, in complete agreement with \cref{eq:initial:condition:for:eta}.

\subsection{Asymptotics of \texorpdfstring{$a$}{a}, \texorpdfstring{$b$}{a} and \texorpdfstring{$\eta$}{eta}}
We are now in the position to evaluate the expressions for $a$, $b$ and $\eta$ in the limit of $\tau_{0} \rightarrow \infty$ using the appropriate asymptotic expansions of the parabolic cylinder functions. For this purpose, we first collect the corresponding expressions, and then apply them to the formulae for $a$, $b$ and $\eta$ derived in the preceding sections of this appendix.
\par
This derivation brings out most clearly that the asymptotic expression for the probability amplitude $a$ relies solely on the elementary expansion of the parabolic cylinder function and the value is given by a logarithmic phase singularity. In contrast, the probability amplitude $b$ requires both the dominant and the subdominant contribution of the parabolic cylinder function. As a result, the derivation of the asymptotic expression for $b$ is more involved than for $a$. Since $\eta$ is the ratio of $a$ and $b$ its asymptotics is mainly governed by $b$.

\subsubsection{Asymptotic expressions for parabolic cylinder functions}
\label{subsubsec:asymptotic:expressions}
In this section, we briefly summarize the asymptotic expressions of the parabolic cylinder functions employed in our analysis. Here, it is crucial to distinguish the behavior of $\mathcal{D}_{\nu}$ at negative times from the one at positive times corresponding to two different phases of $z$.
\par
To bring out this fact most clearly, we first set $z = z_{0}$ in the equations \cref{eq:a:of:z:parabolic:cylinder:function}, \cref{eq:b:of:z:parabolic:cylinder:function} and \cref{eq:representation:eta:parabolic:cylinder:functions}, and choose the scaling 
\begin{align}
    z_0 \equiv \sqrt{2\epsilon}\tau_0\ee^{\ii\frac{\pi}{4}}.
    \label{eq:complex:scaling:z:0}
\end{align}
following from \cref{eq:complex:scaling:appendix} for positive times $\tau_{0}$. 
\par
For negative times we use the representation
\begin{align}
    -z_0 \equiv \ee^{-\ii\pi}z_{0} = \sqrt{2\epsilon}\tau_0\ee^{-\ii\frac{3\pi}{4}}
\end{align}
where we have included the minus sign as a phase using the relation
\begin{align}
    -1 = \ee^{-\ii\pi}.
\end{align}
\par
In the domain \mbox{$\arg[z] < 3\pi/4$} of the complex plane the dominant contribution to the asymptotic expansion of the parabolic cylinder function reads \cite{abramowitz1965, gradshteyn2014} 
\begin{align}
    \parabolicfunction{\rho}{z} \cong \ee^{-\frac{z^{2}}{4}}z^{\rho} 
    \label{eq:asymptotic:expressions:for:prob:2}
\end{align}
which for $\rho = \nu$ and the definition, \cref{eq:definition:of:nu:appendix}, of $\nu$, takes the form
\begin{align}
    \parabolicfunction{\nu}{z} \cong \ee^{-\frac{z^{2}}{4}}z^{-1-\frac{\ii}{2\epsilon}}
    \label{eq:appendix:asymptotic:zero:para:1}
\end{align}
and in the limit $\tau_{0}\rightarrow\infty$ we arrive at
\begin{align}
    \parabolicfunction{\nu}{z} \rightarrow 0.
    \label{eq:appendix:asymptotic:zero:para}
\end{align}
\par
From \cref{eq:asymptotic:expressions:for:prob:2} we find
\begin{align}
    \parabolicfunction{\rho+1}{z} &\cong \ee^{-\frac{z^{2}}{4}}z^{\rho + 1} 
\end{align}
which for $\rho = \nu$ leads us to the expression
\begin{align}
    \begin{split}
        \label{eq:asymptotic:expressions:for:prob:1}
        \parabolicfunction{\nu+1}{z} = \ee^{-\frac{z^{2}}{4}}z^{-\frac{\ii}{2\epsilon}}= \ee^{-\frac{z^{2}}{4}}\ee^{-\frac{\ii}{2\epsilon}\ln\left(z\right)}.
    \end{split}
\end{align}
\par
The two parabolic cylinder functions $\mathcal{D}_{\nu}$ and $\mathcal{D}_{\nu + 1}$ differ by an extra power of $z$ which due to the special form of $\nu$ causes $\mathcal{D}_{\nu}$ to vanish whereas $\mathcal{D}_{\nu + 1}$ contains a logarithmic phase. 
\par
For negative times, that is for the domain \mbox{$-\pi/4>\arg[z]>-5\pi/4$}, the parabolic cylinder function consists \cite{abramowitz1965, gradshteyn2014} to lowest order in $1/z$ of the superposition 
\begin{align}
    \parabolicfunction{\rho}{z} \cong \ee^{-\frac{z^{2}}{4}}z^{\rho }+\frac{\sqrt{2\pi}}{\Gamma(-\rho)}\ee^{-\ii(\rho + 1)\pi}\ee^{\frac{z^{2}}{4}}z^{-\rho-1}.
    \label{eq:asymptotic:expressions:for:prob:4:rho}
\end{align}
\par
Hence, for $\rho = \nu$ only the second term in \cref{eq:asymptotic:expressions:for:prob:4:rho} survives in the asymptotic limit leading us to the expression
\begin{align}
    \parabolicfunction{\nu}{z} \cong \frac{\sqrt{2\pi}}{\Gamma(-\nu)}\ee^{-\ii(\nu + 1)\pi}\ee^{\frac{z^{2}}{4}}z^{-\nu-1}.
    \label{eq:asymptotic:expressions:for:prob:4:nu}
\end{align}
\par
In contrast, for $\rho = \nu + 1$ the first contribution in \cref{eq:asymptotic:expressions:for:prob:4:rho} is the dominant one, and we arrive at
\begin{align}
    \parabolicfunction{\nu+1}{z} \cong \ee^{-\frac{z^{2}}{4}}z^{\nu + 1} = \ee^{-\frac{z^{2}}{4}}z^{-\frac{\ii}{2\epsilon}}.
    \label{eq:asymptotic:expressions:for:prob:3}
\end{align}
We emphasize that for negative times the contributions for $\nu$ and $\nu + 1$ are fundamentally different as expressed by the opposite signs of the Gaussians. Moreover, \cref{eq:asymptotic:expressions:for:prob:4:nu} involves the $\Gamma$ - function which is crucial for the asymptotics of $b$.

\subsubsection{Probability amplitude \texorpdfstring{$a$}{a}}
We now employ the asymptotic expressions derived in the preceding section to obtain the familiar Landau-Zener result for the probability amplitude $a$.
\par
For this purpose, we set in the formula,  \cref{eq:a:of:z:parabolic:cylinder:function} for $a$, $z=z_{0}$, and recall the asymptotic expression, 
\cref{eq:appendix:asymptotic:zero:para} for $\parabolicfunction{\nu}{z_{0}}$ and find 
\begin{align}
    \begin{split}
         \lim_{\tau_{0}\rightarrow\infty}a(\tau_{0}) = \lim_{\tau_{0}\rightarrow\infty}\frac{\parabolicfunction{\nu+1}{-z_{0}
         }}{\parabolicfunction{\nu+1}{z_{0}}}\\
         \label{eq:a:z:0:asymptotic:limit}
    \end{split}
\end{align}
which with \cref{eq:asymptotic:expressions:for:prob:1} and \cref{eq:asymptotic:expressions:for:prob:3} reduces to
\begin{align}
    \begin{split}
         \lim_{\tau_{0}\rightarrow\infty}a(\tau_{0}) = \lim_{\tau_{0}\rightarrow\infty}\frac{\left(-z_{0}\right)^{-\frac{\ii}{2\epsilon}}}{z_{0}^{-\frac{\ii}{2\epsilon}}}.
         \label{eq:a:z:0:asymptotic:limit:step}
    \end{split}
\end{align}
\par
The identity 
\begin{align}
    \left(-1\right)^{-\frac{\ii}{2\epsilon}}=\left(\ee^{-\ii\pi}\right)^{-\frac{\ii}{2\epsilon}}= \ee^{-\frac{\ii}{2\epsilon}\ln\left(\ee^{-\ii\pi}\right)} = \ee^{-\frac{\pi}{2\epsilon}}
\end{align}
leads us to the familiar Landau-Zener result
\begin{align}
    \lim_{\tau_{0}\rightarrow\infty}a(\tau_{0}) = \ee^{-\frac{\pi}{2\epsilon}}.
    \label{eq:asymptotic:expression:a:appendix:alndau:zener}
\end{align}
\par
This derivation brings out three important ingredients of the asymptotic expression for the probability amplitude $a$: (i) It is solely determined by the ratio of of $(-z_{0})^{-\ii/(2\epsilon)}$ and $z_{0}^{-\ii/(2\epsilon)}$. (ii) Since $z_{0}$ cancels there is no need for the limit $\tau_{0}\rightarrow\infty$, and (iii) it is the result of a logarithmic phase singularity $\exp\left[-\ii/\left(2\epsilon\right)\ln\left(-1\right)\right]$ where $-1 = \exp\left(-\ii\pi\right)$. 
\par
Moreover, we emphasize that we have not used the asymptotic expansion, \cref{eq:asymptotic:expressions:for:prob:4:nu}, for $\mathcal{D}_{\nu}$ corresponding to negative times which contains the $\Gamma$ - function. For this reason the derivation of \cref{eq:asymptotic:expression:a:appendix:alndau:zener} is rather straightforward. 

\subsubsection{Probability amplitude \texorpdfstring{$b$}{b}}
Next, we derive the asymptotic expression for the probability amplitude $b$. This analysis is more involved, since in contrast to $a$ which according to \cref{eq:a:z:0:asymptotic:limit} only involves the parabolic cylinder function $\mathcal{D}_{\nu + 1}$ at positive and negative times, we now have to use the parabolic cylinder functions $\mathcal{D}_{\nu}$ and $\mathcal{D}_{\nu + 1}$
for positive and negative times. Hence, we have to employ all four asymptotic expansions obtained in \cref{subsubsec:asymptotic:expressions}. 
\par
From \cref{eq:b:of:z:parabolic:cylinder:function} we find with the help of the asymptotic formula, \cref{eq:appendix:asymptotic:zero:para}, the expression
\begin{widetext}
\begin{align}
    \begin{split}
         \lim_{\tau_{0}\rightarrow\infty}b(\tau_{0}) =\lim_{\tau_{0}\rightarrow\infty}\frac{-\sqrt{2\epsilon}\ee^{-\ii\frac{\pi}{4}}}{\parabolicfunction{\nu}{-z_{0}}}\left[\parabolicfunction{\nu+1}{z_{0}} - \frac{\parabolicfunctionsquare{\nu+1}{-z_{0}}}{\parabolicfunction{\nu+1}{z_{0}}}\right]
        \label{eq:b:z:0:asymptotic:limit}
    \end{split}
\end{align}
which due to \cref{eq:asymptotic:expressions:for:prob:1}, \cref{eq:asymptotic:expressions:for:prob:4:nu} and \cref{eq:asymptotic:expressions:for:prob:3} takes the form
\begin{align}
    \begin{split}
         \lim_{\tau_{0}\rightarrow\infty}b(\tau_{0}) =\lim_{\tau_{0}\rightarrow\infty}\frac{-\sqrt{2\epsilon}\ee^{-\ii\frac{\pi}{4}}}{\frac{\sqrt{2\pi}}{\Gamma(-\nu)}\ee^{-\ii(\nu + 1)\pi}\ee^{\frac{z_{0}^{2}}{4}}(-z_{0})^{-\nu-1}}\left[\ee^{-\frac{z_{0}^{2}}{4}}z_{0}^{\nu + 1} - \frac{\left(\ee^{-\frac{z_{0}^{2}}{4}}(-z_{0})^{\nu + 1}\right)^{2}}{\ee^{-\frac{z_{0}^{2}}{4}}z_{0}^{\nu + 1}}\right]
        \label{eq:b:z:0:asymptotic:limit:additional}
    \end{split}
\end{align}
or
\begin{align}
    \lim_{\tau_{0}\rightarrow\infty}b(\tau_{0}) = \lim_{\tau_{0}\rightarrow\infty}\sqrt{\frac{\epsilon}{\pi}}\Gamma(-\nu)\ee^{-\ii\frac{5\pi}{4}}\frac{\ee^{-\frac{z_{0}^{2}}{2}}z_{0}^{2(\nu +1)}}{\ee^{-\ii(\nu + 1)\pi}(-1)^{-\nu-1}}\left[1-\left(-1\right)^{2(\nu+1)}\right].
\end{align}
\end{widetext}
\par
With the identities
\begin{align}
    \ee^{-\frac{z_{0}^{2}}{2}} = \ee^{-\ii\epsilon\tau_{0}^{2}}
    \label{eq:appendix:complex:gaussian}
\end{align}
and
\begin{align}
    z_{0}^{2(\nu +1)} = \ee^{-\frac{\ii}{\epsilon}\ln\left(z_{0}\right)} = \ee^{-\frac{\ii}{\epsilon}\ln\left(\sqrt{2\epsilon}\tau_{0}\right)}\ee^{\frac{\pi}{4\epsilon}}
    \label{eq:appendix:logarithmic:phase}
\end{align}
following from the scaling, \cref{eq:complex:scaling:z:0}, of $\tau_{0}$ together with the relation 
\begin{align}
    \ee^{-\ii(\nu + 1)\pi}(-1)^{-\nu-1} = 1
\end{align}
we arrive at the expression
\begin{align}
    b(\tau_{0}) = \sqrt{\frac{\epsilon}{\pi}}\Gamma(-\nu)\ee^{\ii\xi}\ee^{\frac{\pi}{4\epsilon}}\left[1-\ee^{-\frac{\pi}{\epsilon}}\right]
    \label{eq:appendix:b:asymptotic:limit}
\end{align}
where we have introduced the phase
\begin{align}
    \xi \equiv -\frac{5\pi}{4}-\epsilon\tau_{0}^{2}-\frac{1}{\epsilon}\ln(\sqrt{2\epsilon}\tau_{0}).
\end{align}
\par
Next, we recall the definition, \cref{eq:definition:of:nu:appendix}, of $\nu$, and use the decomposition
\begin{align}
    \Gamma(-\nu) = \abs{\Gamma\left(1+\frac{\ii}{2\epsilon}\right)}\ee^{\ii\arg\left[\Gamma\left(1+\frac{\ii}{2\epsilon}\right)\right]}
    \label{eq:appendix:decomposition:gamma:function}
\end{align}
into amplitude and phase, together with the functional equation \cite{abramowitz1965, gradshteyn2014}
\begin{align}
    \Gamma\left(1+\frac{\ii}{2\epsilon}\right) = \frac{\ii}{2\epsilon}\Gamma\left(\frac{\ii}{2\epsilon}\right)
\end{align}
and the identity \cite{abramowitz1965, gradshteyn2014}
\begin{align}
    \abs{\Gamma\left(\frac{\ii}{2\epsilon}\right)}&=\sqrt{\frac{\pi}{\frac{1}{2\epsilon}\sinh\left(\frac{\pi}{2\epsilon}\right)}} = 2\sqrt{\pi\epsilon}\frac{\ee^{-\frac{\pi}{4\epsilon}}}{\sqrt{1-\ee^{-\frac{\pi}{\epsilon}}}}
    \label{eq:appendix:identity:gamma:function:abs:sinh}
\end{align}
of the $\Gamma$-function to further simplify the expression, \cref{eq:appendix:b:asymptotic:limit}, for $b$.
\par
After minor algebra we arrive at the final expression
\begin{align}
    b(\tau_{0}) = \sqrt{1-\ee^{-\frac{\pi}{\epsilon}}}\ee^{\ii\phi}
    \label{eq:asymptotic:expression:b:appendix}
\end{align}
where we have introduced the phase
\begin{align}
    \begin{split}
        \phi \equiv -\frac{5\pi}{4}&-\epsilon\tau_{0}^2-\frac{1}{\epsilon}\ln\left(\sqrt{2\epsilon}\tau_{0}\right)\\ &+ \arg\left[\Gamma\left(1+\frac{\ii}{2\epsilon}\right)\right].
    \end{split}
\end{align}
\par
It is interesting to identify the origins of the amplitude of $b$ and the individual contributions to the phase $\phi$. Indeed, the square root in \cref{eq:asymptotic:expression:b:appendix} is a consequence of the difference of the parabolic cylinder functions in \cref{eq:b:z:0:asymptotic:limit} and the appearance of the $\Gamma$ - function in \cref{eq:asymptotic:expressions:for:prob:4:nu} whose absolute value reduces with the help of \cref{eq:appendix:identity:gamma:function:abs:sinh} the expression in the square brackets in \cref{eq:appendix:b:asymptotic:limit} to the square root.
\par
According to \cref{eq:appendix:b:asymptotic:limit} and \cref{eq:appendix:decomposition:gamma:function} the phase of the $\Gamma$ - function is one contribution to phase $\phi$. The others arise from the complex-valued Gaussian, \cref{eq:appendix:complex:gaussian}, which appears in the asymptotic expansions, \cref{eq:asymptotic:expressions:for:prob:1} and \cref{eq:asymptotic:expressions:for:prob:4:nu}, of $\parabolicfunction{\nu+1}{z_{0}}$ and $\parabolicfunction{\nu}{-z_{0}}$ with opposite signs leading to a factor of two.
\par
The logarithmic phase emerges in \cref{eq:appendix:logarithmic:phase}, since the parabolic cylinder functions $\parabolicfunction{\nu+1}{z_{0}}$ and $\parabolicfunction{\nu+1}{-z_{0}}$ given by 
\cref{eq:asymptotic:expressions:for:prob:1} and \cref{eq:asymptotic:expressions:for:prob:3} are proportional to $z^{\nu + 1}$ whereas, according to \cref{eq:asymptotic:expressions:for:prob:4:nu}, $\parabolicfunction{\nu}{-z_{0}}$ is proportional to $z^{-\nu - 1 }$.
\par
The constant phase of $-5\pi/4$ is a consequence of the minus sign in the definition, \cref{eq:b:of:z:parabolic:cylinder:function},  of $b$ and the transformation, \cref{eq:complex:scaling:appendix}, from $\tau$ to $z$.

\subsubsection{Solution \texorpdfstring{$\eta$}{eta} of the Riccati equation}
Finally, we address the asymptotic behavior of the solution $\eta$ of the Riccati equation, \cref{eq:riccati:equation:eta}. With the help of the asymptotic expressions, \cref{eq:asymptotic:expression:a:appendix:alndau:zener}
and \cref{eq:asymptotic:expression:b:appendix}, for the probability amplitudes $a$ and $b$, we find from \cref{eq:appendix:eta:b:over:a} the result
\begin{align}
    \eta(\tau_{0}) = \frac{\sqrt{1-\ee^{-\frac{\pi}{\epsilon}}}}{\ee^{-\frac{\pi}{2\epsilon}}}\ee^{\ii\phi_{\eta}(\tau_{0})}
\end{align}
where we have defined the phase
\begin{align}
    \begin{split}
        \phi_{\eta} \equiv -\frac{\pi}{4}&-\epsilon\tau_{0}^2-\frac{1}{\epsilon}\ln\left(\sqrt{2\epsilon}\tau_{0}\right)\\ &+ \arg\left[\Gamma\left(1+\frac{\ii}{2\epsilon}\right)\right].
    \end{split}
\end{align}
of $\eta$. 
\par
Needless to say, the same result follows from \cref{eq:representation:eta:parabolic:cylinder:functions} when we apply the asymptotic expressions, \cref{eq:appendix:asymptotic:zero:para}, \cref{eq:asymptotic:expressions:for:prob:1}, \cref{eq:asymptotic:expressions:for:prob:4:nu} and \cref{eq:asymptotic:expressions:for:prob:3}. 

\section{Origin of oscillations at early times}
\label{appendix:B}
In \cref{section:5}, we have identified oscillations in an approximate analytical expression for $\eta_{M}$, valid for large negative times. We dedicate this appendix to demonstrating that they result from the superposition of the two instantaneous eigenstates \cite{Suominen1991, Kocharovsky1997, ivakhnenko2023, Steck2024} of the time-dependent Hamiltonian, \cref{eq:Hamiltonian}.
\par
In order to bring out this fact most clearly, we first briefly review essential ingredients of this approach and then employ it to derive an expression for the amplitude $a$ valid for large negative times. Finally, we compare and contrast the so-obtained approximation for $a$ to \cref{eq:asymptotic:expression:for:H:M:large:negative:times} following from the Markov solution, \cref{eq:a:M:negative:times}.

\subsection{Instantaneous eigenstates}
We start by recalling the expressions of the instantaneous eigenvalues and eigenstates. For this purpose, we cast the equations, \cref{eq:diff:equation:a} and \cref{eq:diff:equation:b}, for the probability amplitudes $a$ and $b$ into the matrix form 
\begin{align}
    \ii\dv{\tau}
    \begin{pmatrix}
        a(\tau) \\
        b(\tau)
    \end{pmatrix} 
    = 
    \begin{pmatrix}
        -\epsilon\tau & 1 \\
        1 & \epsilon\tau
    \end{pmatrix}
    \begin{pmatrix}
        a(\tau) \\
        b(\tau)
    \end{pmatrix}
    \label{eq:schrödinger:equation:diagonailzation}
\end{align}
which leads us to the instantaneous eigenvalues 
\begin{align}
    \theta_{\pm}(\tau) \equiv \pm\theta(\tau) \equiv \pm\sqrt{\left(\epsilon\tau\right)^{2} + 1}
    \label{eq:theta:appendix}
\end{align}
with the corresponding normalized eigenstates
\begin{align}
    u_{\pm}(\tau) \equiv N_{\pm}(\tau)  
    \begin{pmatrix}
        v_{\pm}(\tau) \\
        1
    \end{pmatrix}.
    \label{eq:instanteneous:eigenstates:u:plusminus}
\end{align}
\par
Here, we have introduced the abbreviation
\begin{align}
    v_{\pm}(\tau) \equiv -\epsilon\tau \pm \theta
    \label{eq:v:mp:appendix}
\end{align}
and the normalization constant
\begin{align}
    N_{\pm}(\tau) \equiv \frac{1}{\sqrt{1+v^{2}_{\pm}(\tau)}}.
\end{align}
\par
In order to connect the diabatic with the adiabatic basis we consider the limit of $\tau\rightarrow\infty$. From the definitions, \cref{eq:theta:appendix} and \cref{eq:v:mp:appendix} we find 
\begin{align}
    v_{+}(\tau) \cong
    \begin{cases}
        2\epsilon\abs{\tau} & \tau<0 \\
        \frac{1}{2\epsilon\abs{\tau}} & \, \tau>0
    \end{cases}
\end{align}
and
\begin{align}
    v_{-}(\tau) \cong 
    \begin{cases}
        -\frac{1}{2\epsilon\abs{\tau}} & \tau<0 \\
        -2\epsilon\abs{\tau} & \, \tau>0
    \end{cases}
\end{align}
which leads to the asymptotic eigenstates
\begin{align}
    u_{-} \cong 
    \begin{pmatrix}
        0 \\
        1
    \end{pmatrix}
\end{align}
and
\begin{align}
    u_{+} \cong 
    \begin{pmatrix}
        1 \\
        0
    \end{pmatrix}
\end{align}
for large \textit{negative} times, whereas for large \textit{positive} times we find 
\begin{align}
    u_{-} \cong 
    \begin{pmatrix}
        -1 \\
        0
    \end{pmatrix}
\end{align}
and
\begin{align}
    u_{+} \cong 
    \begin{pmatrix}
        0 \\
        1
    \end{pmatrix}.
\end{align}
\par
Hence, $u_{+}$ and $u_{-}$ correspond asymptotically to the diabatic states $a$ and $b$.

\subsection{Diagonalization}
Next, we diagonalize the Hamiltonian, \cref{eq:Hamiltonian}, using the instantaneous eigenstates, \cref{eq:instanteneous:eigenstates:u:plusminus}. Needless to say, this technique is only correct under appropriate conditions. In particular, we show it is valid for large negative times which represents the domain of interest in the present discussion.
\par
For the diagonalization we now find the transformation matrix 
\begin{align}
    \begin{split}
    G &\equiv   
        \begin{pmatrix}
        G_{11} & G_{12}  \\
        G_{21} & G_{22} 
    \end{pmatrix}
    \\
    &=
    \begin{pmatrix}
        \frac{v_{+}}{\sqrt{1+v_{+}^{2}}} & \frac{v_{-}}{\sqrt{1+v_{-}^{2}}}  \\
        \frac{1}{\sqrt{1+v_{+}^{2}}} & \frac{1}{\sqrt{1+v_{-}^{2}}} 
    \end{pmatrix}
    \label{eq:G:appendix}
     \end{split}
\end{align}
from the diabatic to the adiabatic basis, and the corresponding inverse matrix
\begin{align}
    \begin{split}
    G^{-1} &\equiv 
    \begin{pmatrix}
        G^{-1}_{11} & G^{-1}_{12}  \\
        G^{-1}_{21} & G^{-1}_{22} 
    \end{pmatrix}
    \\
    &=
    \begin{pmatrix}
        \frac{\sqrt{1+v_{+}^{2}}}{v_{+}-v_{-}} & -\frac{\sqrt{1+v_{+}^{2}}v_{-}}{v_{+}-v_{-}}\\
        -\frac{\sqrt{1+v_{-}^{2}}}{v_{+}-v_{-}} & \frac{\sqrt{1+v_{-}^{2}}v_{+}}{v_{+}-v_{-}}
    \end{pmatrix}
    \label{eq:G:minus:1:appendix}
    \end{split}
\end{align}
which is the transformation from the adiabatic to the diabatic basis, that is
\begin{align}
    \begin{pmatrix}
        a(\tau) \\
        b(\tau)
    \end{pmatrix} 
    = G(\tau)    
    \begin{pmatrix}
        \tilde{a}(\tau)  \\
        \tilde{b}(\tau)
    \end{pmatrix}.
    \label{eq:transformation:adiabatic:to:diabatic:basis}
\end{align}
\par
The matrices $G$ and $G^{-1}$ diagonalize the Hamiltonian $\mathcal{H}$
and the Schrödinger equation, \cref{eq:schrödinger:equation:diagonailzation}, takes the form
\begin{align}
    \ii\dv{\tau}
    \begin{pmatrix}
        \tilde{a} \\
        \tilde{b}
    \end{pmatrix} 
    = D
    \begin{pmatrix}
        \tilde{a} \\
        \tilde{b}
    \end{pmatrix}
    - \ii G^{-1}\left(\dv{\tau}G\right)
    \begin{pmatrix}
        \tilde{a} \\
        \tilde{b}
    \end{pmatrix}
    \label{eq:schrödinger:equation:adiabatic:basis:appendix}
\end{align}
where 
\begin{align}
    D \equiv G^{-1}    
    \begin{pmatrix}
        -\epsilon\tau & 1 \\
        1 & \epsilon\tau
    \end{pmatrix}
    G.
\end{align}
\par
With the help of the explicit expressions, \cref{eq:G:appendix} and \cref{eq:G:minus:1:appendix}, for $G$ and $G^{-1}$, it is easy to verify that $\mathcal{D}$ is indeed the diagonal matrix
\begin{align}
    D(\tau) = 
    \begin{pmatrix}
        \theta(\tau) & 0  \\
        0  & -\theta(\tau)
    \end{pmatrix}
\end{align}
where $\theta$ is defined by \cref{eq:theta:appendix}.

\subsection{Solution for the diagonalized system}
We emphasize that we can solve \cref{eq:schrödinger:equation:adiabatic:basis:appendix} only when we can neglect the time derivative of $G$ governing the last term. Since in the limit of large negative times the matrix 
\begin{align}
    G(-\abs{\tau}) \cong 
    \begin{pmatrix}
       1 & -\frac{1}{2\epsilon\abs{\tau}}\\
       \frac{1}{2\epsilon\abs{\tau}} & 1
    \end{pmatrix}
    \label{eq:G:asymptotically:appendix}
\end{align}
is almost constant we can indeed neglect it.
\par
Therefore, we solve the equation 
\begin{align}
     \ii\dv{\tau}
    \begin{pmatrix}
        \tilde{a} \\
        \tilde{b}
    \end{pmatrix} 
    = D
    \begin{pmatrix}
        \tilde{a} \\
        \tilde{b}
    \end{pmatrix}   
\end{align}
obtained from \cref{eq:schrödinger:equation:adiabatic:basis:appendix} in the limit of large negative times, and find the solutions 
\begin{align}
    \tilde{a}(-\abs{\tau})) = \tilde{a}(-\tau_{0})\ee^{-\ii\chi(-\abs{\tau})}
    \label{eq:a:tilde:adiabatic:basis}
\end{align}
and
\begin{align}
    \tilde{b}(-\abs{\tau})) = \tilde{b}(-\tau_{0})\ee^{\ii\chi(-\abs{\tau})}.
    \label{eq:b:tilde:adiabatic:basis}
\end{align}
\par
Here, we have introduced the time-dependent function 
\begin{align}
    \chi(\tau) \equiv \int\limits_{-\tau_{0}}^{\tau}\dd\tau^{\prime}\theta(\tau^{\prime}).
    \label{eq:appendix:definition:chi}
\end{align}
\par
The initial conditions $\tilde{a}(-\tau_{0})$ and $\tilde{b}(-\tau_{0})$ in \cref{eq:a:tilde:adiabatic:basis} and \cref{eq:b:tilde:adiabatic:basis} are connected to the initial conditions $a(-\tau_{0})$ and $b(-\tau_{0})$, \cref{eq:initial:condition:a} and \cref{eq:initial:condition:b}, by the matrix $G^{-1}$, that is 
\begin{align}
    \begin{pmatrix}
        \tilde{a}(\tau) \\
        \tilde{b}(\tau)
    \end{pmatrix} 
    = G^{-1}(\tau)    
    \begin{pmatrix}
        a(\tau)  \\
        b(\tau)
    \end{pmatrix}
\end{align}
which with the choice of the initial conditions, \cref{eq:initial:condition:a} and \cref{eq:initial:condition:b}, reduces to
\begin{align}
    \tilde{a}(-\tau_{0}) = G^{-1}_{11}(-\tau_{0})
    \label{eq:initial:condition:a:tilde}
\end{align}
and 
\begin{align}
    \tilde{b}(-\tau_{0}) = G^{-1}_{21}(-\tau_{0}).
    \label{eq:initial:condition:b:tilde}
\end{align}
\par
Now we are in the position to obtain an expression for the probability amplitude $a$ in the diabatic basis. For this purpose, we substitute \cref{eq:G:appendix} into \cref{eq:transformation:adiabatic:to:diabatic:basis} and find 
\begin{align}
    a(-\abs{\tau}) = G_{11}(-\abs{\tau})\tilde{a}(-\abs{\tau}) + G_{12}(-\abs{\tau})\tilde{b}(-\abs{\tau})
\end{align}
which with the help of the solutions, \cref{eq:a:tilde:adiabatic:basis} and \cref{eq:b:tilde:adiabatic:basis}, in the adiabatic basis, and the initial conditions, \cref{eq:initial:condition:a:tilde} and \cref{eq:initial:condition:b:tilde}, takes the form
\begin{align}
    \begin{split}
        a(-\abs{\tau})) &= G_{11}(-\abs{\tau})G^{-1}(-\tau_{0})\ee^{-\ii\chi(-\abs{\tau})}\\
        &+ G_{12}(-\abs{\tau})G^{-1}_{21}(-\tau_{0})\ee^{\ii\chi(-\abs{\tau})}.
        \label{eq:appendix:formal:solution:for:a}
    \end{split}
\end{align}
\par
We emphasize apart from the assumption that $G$ is constant for large negative times, we have made no further approximations.

\subsection{Asymptotic behavior for large negative times}

Since we are interested in an approximate expression for $a$ in the neighborhood of $\tau=-\tau_{0}$ we expand $\theta$ and $v_{\pm}$ for large negative times, and find the expressions 
\begin{align}
    \theta_{\pm}(-\abs{\tau}) \cong \pm\left(\epsilon\abs{\tau} + \frac{1}{2\epsilon\abs{\tau}}\right)
    \label{eq:appendix:asymptotic:behavoir:theta}
\end{align}
as well as
\begin{align}
     v_{+}(-\abs{\tau}) \cong 2\epsilon\abs{\tau}
\end{align}
and 
\begin{align}
    v_{-}(-\abs{\tau}) \cong -\frac{1}{2\epsilon\abs{\tau}}.
\end{align}
\par
Hence, the difference of $v_{+}$ and $v_{-}$ is given by
\begin{align}
    v_{+} - v_{-} = 2\epsilon\abs{\tau} + \frac{1}{2\epsilon\abs{\tau}}. 
\end{align}
\par
Now we are in the position to approximate each matrix element in \cref{eq:appendix:formal:solution:for:a}, we find 
\begin{align}
    G_{11}(-\abs{\tau}) = \frac{v_{+}}{\sqrt{1+v_{+}^{2}}} \cong 1
    \label{eq:appendix:approximate:expression:for:G:11}
\end{align}
and 
\begin{align}
    G_{12}(-\abs{\tau}) = \frac{v_{-}}{\sqrt{1+v_{-}^{2}}} \cong -\frac{1}{2\epsilon\abs{\tau}}.
\end{align}
\par
Moreover, we find the expressions
\begin{align}
    G_{11}^{-1}(-\tau_{0}) = \frac{\sqrt{1+v_{+}(-\tau_{0})^{2}}}{v_{+}(-\tau_{0})-v_{-}(-\tau_{0})} \cong 1 - \frac{1}{4\epsilon^{2}\tau_{0}^{2}}
\end{align}
and 
\begin{align}
    G_{21}^{-1}(-\tau_{0}) = -\frac{\sqrt{1+v_{-}(-\tau_{0})^{2}}}{v_{+}(-\tau_{0})-v_{-}(-\tau_{0})} \cong -\frac{1}{2\epsilon\abs{\tau_{0}}}
    \label{eq:appendix:approximate:expression:for:G:21:minus:tau:0}
\end{align}
for the matrix elements of the inverse matrix $G^{-1}$, dependent on the initial time $-\tau_{0}$.
\par
Next, we insert \cref{eq:appendix:approximate:expression:for:G:11} to \cref{eq:appendix:approximate:expression:for:G:21:minus:tau:0} into \cref{eq:appendix:formal:solution:for:a} which yields
\begin{align}
    \begin{split}
        a(-\abs{\tau}) = \left(1 - \frac{1}{4\epsilon^{2}\tau_{0}^{2}}\right)\ee^{-\ii\chi(-\abs{\tau})}
        +\frac{1}{4\epsilon^{2}\tau_{0}\abs{\tau}}\ee^{\ii\chi(-\abs{\tau})}
    \end{split}
\end{align}
where we do not yet use the asymptotic expression for $\theta$. 
\par
In the next step, we factor out the first exponential and obtain the expression
\begin{align}
    a(-\abs{\tau}) = \ee^{-\ii\chi(-\abs{\tau})}\left(1 - \frac{1}{4\epsilon^{2}\tau_{0}^{2}}+\frac{1}{4\epsilon^{2}\tau_{0}\abs{\tau}}\ee^{2\ii\chi(-\abs{\tau})}\right)
\end{align}
which with the help of the Taylor expansion
\begin{align}
    \ee^{X} \cong 1 + X
\end{align}
of the exponential function yields 
\begin{align}
    a(-\abs{\tau}) \cong \ee^{-\ii\chi(-\abs{\tau})}\exp\left(-\frac{1}{4\epsilon^{2}\tau_{0}^{2}}+\frac{1}{4\epsilon^{2}\tau_{0}\abs{\tau}}\ee^{2\ii\chi(-\abs{\tau})}\right).
    \label{eq:appendix:expression:for:a:with:chi}
\end{align}
\par
Next, we evaluate the function $\chi$, \cref{eq:appendix:definition:chi}, for large negative times by inserting the asymptotic behavior of $\theta$, \cref{eq:appendix:asymptotic:behavoir:theta}, which leads us to
\begin{align}
    \chi(-\abs{\tau}) \cong \frac{\epsilon}{2}\left(\tau_{0}^2-\tau^{2}\right) + \frac{1}{2\epsilon}\ln\left(\frac{\abs{\tau}}{\tau_{0}}\right).
    \label{eq:appendix:expression:for:chi:large:negative:times}
\end{align}
\par
Thus, we arrive at the expression
\begin{align}
    \begin{split}
        a_{M}(-\abs{\tau})=\ee^{-\ii\epsilon\left(\tau_{0}^{2}-\tau^{2}\right)/2}\ee^{-\bar{\mathrm{H}}_{M}(-\abs{\tau})}
        \label{eq:a:M:negative:times:appendix:B}
    \end{split}
\end{align}
where 
\begin{align}
    \bar{\mathrm{H}}_{M}(-\abs{\tau}) \equiv \frac{\ii}{2\epsilon}\ln\left(\frac{\abs{\tau}}{\tau_{0}}\right)+\frac{1}{4\epsilon^{2}\tau_{0}^{2}} - \frac{\ee^{\ii\epsilon\left(\tau_{0}^{2}-\tau^{2}\right)} \ee^{\frac{\ii}{\epsilon}\ln\left(\frac{\abs{\tau}}{\tau_{0}}\right)}}{4\epsilon^{2}\tau_{0}\abs{\tau}}.
    \label{eq:appendix:B:final:result}
\end{align}
Here we have used \cref{eq:appendix:expression:for:chi:large:negative:times} and \cref{eq:appendix:expression:for:a:with:chi}. 
\par
On first sight, this expression seems to be different from \cref{eq:a:M:negative:times} due to the presence of the logarithmic phase singularity in the exponential function. However, the asymptotic expansion
\begin{align}
    \ln\left(\frac{\abs{\tau}}{\tau_{0}}\right) = \ln\left[1+\frac{\abs{\tau}-\tau_{0}}{\tau_{0}}\right] \cong \frac{\abs{\tau}-\tau_{0}}{\tau_{0}}
\end{align}
of the logarithm shows that the correction scales like $1/\tau_{0}$ and leads to a higher order correction.
\par
For this reason, we find 
\begin{align}
    \mathrm{H}_{M}(-\abs{\tau}) \cong \bar{\mathrm{H}}_{M}(-\abs{\tau}),
\end{align}
and \cref{eq:appendix:B:final:result} reduces indeed to \cref{eq:asymptotic:expression:for:H:M:large:negative:times}. 
                                        
\nocite{*}

\bibliography{references}

\end{document}